\documentclass[longauth]{aa}  

\usepackage{graphicx}
\usepackage{gensymb}

\usepackage{txfonts}
\usepackage[flushleft]{threeparttable}
\renewcommand{\arraystretch}{1.1}
\usepackage[hidelinks,colorlinks=true,linkcolor=blue,citecolor=blue]{hyperref}
\usepackage{xcolor}
\usepackage[section]{placeins}
\usepackage{longtable}
\usepackage{lipsum}
\usepackage{adjustbox}
\usepackage{stfloats}
\usepackage{scrextend}
\usepackage{amssymb}
\usepackage{amsmath}
\usepackage{mathrsfs}
\usepackage{multicol}
\usepackage{booktabs}
\usepackage{moresize}
\usepackage{orcidlink}
\usepackage{xspace}
\usepackage{placeins}
\usepackage{caption}

\newcommand{\kms}{\mbox{km\,s$^{-1}$}\xspace}
\newcommand{\ms}{\mbox{m\,s$^{-1}$}\xspace}
\newcommand{\msyr}{\mbox{m\,s$^{-1}$\,yr$^{-1}$}\xspace}
\newcommand{\msyryr}{\mbox{m\,s$^{-1}$\,yr$^{-2}$}\xspace}

\newcommand{\mjup}{\mbox{$\it{M_{\rm \mathrm{Jup}}}$}\xspace}

\newcommand{\msol}{\mbox{$\it{M_\mathrm{\odot}}$}\xspace}

\begin{document}

   \title{Companions of Hot Jupiters: two decades of radial velocity monitoring with CORALIE}

\titlerunning{}
\authorrunning{L. Parc, et al.}
   \author{L\'ena~Parc\inst{1,*}, 
    Jo\~ao P.~Faria\inst{1},
    Fran\c{c}ois~Bouchy\inst{1},
    Jana~Baron\inst{1,2},
    Siddharth~Bhatnagar\inst{1,3,4},
    Nicolas~Billot\inst{1},
    Yann~Carteret\inst{1},
    Guillaume~Chaverot\inst{1,5},
    Adrien~Deline\inst{1},
    Jean-Baptiste~Delisle\inst{1},
    Pedro~Figueira\inst{1,6},
    Emile~Fontanet\inst{1},
    Nathan~C.\,Hara\inst{1,2},
    Mathilde~Houelle\inst{1},
    Maxime~Marmier\inst{1},
    Marion~Neveu-VanMalle\inst{1},
    Louise~D.\,Nielsen\inst{1,7},
    Javiera~Rey\inst{1},
    Damien~Ségransan\inst{1},
    Manu~Stalport\inst{1,8,9},
    Stéphane~Udry\inst{1}
          }

   \institute{
       \inst{1} Observatoire de Gen\`eve, D\'epartement d’Astronomie, Universit\'e de Gen\`eve, Chemin Pegasi 51, 1290 Versoix, Switzerland\\
       \inst{2} Aix Marseille Université, CNRS, CNES, LAM, Marseille, France\\
       \inst{3} Centre pour la vie dans l'Univers de l'Université de Genève, Genève, Switzerland\\
        \inst{4} Group of Applied Physics and Institute for Environmental Sciences, University of Geneva, Geneva, Switzerland\\
        \inst{5} Univ. Grenoble Alpes, CNRS, IPAG, 38000 Grenoble, France\\
        \inst{6} Instituto de Astrof\'{i}sica de Andaluc\'{i}a-CSIC, Glorieta de la Astronom\'{i}a s/n, E-18008 Granada, Spain\\
        \inst{7} University Observatory, Faculty of Physics, Ludwig-Maximilians-Universit{\"a}t M{\"u}nchen, Scheinerstr. 1, 81679 Munich, Germany\\
        \inst{8} Space Sciences, Technologies and Astrophysics Research (STAR) Institute, Université de Liège, Allée du 6 Août 19C, 4000 Liège, Belgium\\
        \inst{9} Astrobiology Research Unit, Université de Liège, Allée du 6 Août 19C, B-4000 Liège, Belgium\\
        \inst{*}\email{lena.parc@unige.ch}    
             }

   \date{Received 31 August 2026; accepted 14 October 2026}

 
    \abstract{Hot Jupiters are thought to form through multiple pathways, some of which can leave observable signatures in the architecture of their planetary systems. Massive outer companions, in particular, can provide important constraints on the dynamical histories of hot Jupiters.}{We aim to characterize the occurrence rate and orbital properties of additional giant companions in hot Jupiter systems, and to investigate whether these companions provide evidence for the role of dynamical interactions in shaping the hot Jupiter population.}{We perform a homogeneous long-term radial-velocity analysis of 147 WASP hot Jupiter systems observed with the CORALIE spectrograph between 2007 and 2025. We identify or refine 13 outer giant companions and use individual radial-velocity posteriors within a Bayesian occurrence-rate framework to measure the frequency of giant companions over orbital periods of 10--6250 days.}{The occurrence rate of additional giant companions increases strongly with orbital period, from $<0.01$ and $0.02\pm0.01$ at 10--50 and 50--250 days to $0.05\pm0.03$ and $0.17\pm0.06$ at 250--1250 and 1250--6250 days, respectively, in agreement with the giant-planet occurrence of the general FGK population and with previous estimates of outer companions to hot Jupiters. Hot Jupiters with detected companions have longer orbital periods and higher eccentricities than those without. We also find that companion eccentricity decreases with hot Jupiter period, and that the companion-to-hot-Jupiter mass ratio increases with period ratio, both consistent with expectations from high-eccentricity migration (HEM).}{Our results support HEM as the mechanism shaping hot Jupiters with outer companions. Combining long-baseline radial velocities with future \textit{Gaia} astrometry will further constrain companion masses, mutual inclinations, and dynamical histories.}


   \maketitle
%

\section{Introduction}

Hot Jupiters (H-Js) were the first exoplanets to be discovered around main sequence stars and surprised us with their unexpectedly close-in orbits \citep{Mayor1995}. They are a prime example of how exoplanets have challenged our knowledge of how planetary systems form and evolve. More than thirty years after the discovery of the first H-J, there is no consensus on their predominant origin channel. 

Three formation scenarios have been proposed. In situ formation requires unusually massive protoplanetary disks to build a giant planet so close to its host star \citep{Bodenheimer2000,Batygin2016}. Disk-driven migration instead assumes the planet forms beyond the ice line before being shepherded inward through interactions with the gas disk \citep{Lin1996,Ward1997}. Finally, high-eccentricity tidal migration invokes dynamical excitation through planet-planet scattering \citep{Rasio1996,Chatterjee2008}, Lidov-Kozai cycling induced by a distant stellar or planetary companion \citep{Wu2003,Naoz2012,Petrovich2016}, or secular interactions \citep{Wu2011,Petrovich2015}, followed by tidal circularization onto a tight orbit. These mechanisms predict different system architectures: disk migration tends to produce well-aligned circular orbits, while dynamical pathways more naturally explain eccentric and misaligned systems.

Observational tests have yielded mixed conclusions. Distributions of orbital parameters and host-star obliquities broadly favor high-eccentricity migration as a major contributor to the population \citep{Fabrycky2009,Bonomo2017,Rice2022}. However, direct imaging surveys find that fewer than 20\% of H-J systems host a stellar companion close enough to drive Kozai-Lidov oscillations \citep{Ngo2016}, and the majority of Kepler H-Js show no evidence for nearby planetary companions \citep{Steffen2012}. Long-baseline radial-velocity surveys find good agreement on the presence of massive outer companions to H-Js \citep{Knutson2014, Bryan2016,Zink2023}. However, these studies rest on limited observational constraints: the occurrence rates from \citet{Knutson2014} and \citet{Bryan2016} are dominated by long-term radial velocity (RV) trends with partially resolved orbits rather than fully characterized companions, while the \citet{Zink2023} estimate is based on only 5 detected companions among 11 H-Js hosts. In addition, some individual systems show clear signatures of high-eccentricity migration \citep{Wu2003,Gupta2024}, while others have architectures that are impossible to reconcile with it \citep{Becker2015,Hord2022}.

No single formation channel can explain all observed H-J properties, and the most plausible picture is that disk-driven and high-eccentricity migration are complementary rather than competing channels, together accounting for the bulk of the population \citep{Anderson2015}. Identifying and characterizing companions in H-J systems remains one of the most direct ways to constrain their formation history, motivating dedicated surveys such as the \textit{Friends of Hot Jupiters program} at Keck/HIRES \citep{Knutson2014,Ngo2015,Ngo2016} and the long-term CORALIE follow-up of WASP systems presented in this work, whose large sample size and nearly two-decade single-instrument baseline offer good control over systematic biases and sensitivity to multi-year companion orbits.

In this paper, we present a homogeneous radial velocity analysis of 147 WASP H-J systems observed with CORALIE between 2007 and end of 2025. Section~\ref{sect:observations} describes the survey, the target selection, and the data reduction. Section~\ref{sect:rv_analysis} presents the radial velocity modeling framework and the detected companions. Section~\ref{sect:occurrence_rates} derives the occurrence rate of companions to H-Js as a function of orbital period. Section~\ref{sect:discussion} discusses the implications of these results for H-J formation, and Section~\ref{sect:conclusions} summarizes our conclusions.


\section{The CORALIE long-term follow-up of the WASP hot Jupiters}\label{sect:observations}

\subsection{Description of the survey}

The CORALIE spectrograph \citep{Queloz2000}, mounted on the 1.2m Euler Swiss Telescope at La Silla Observatory, has been used to follow up southern WASP H-J systems \citep{Pollacco2006} for nearly two decades. Once a planet was confirmed through a small number of radial velocity measurements, the search for additional outer companions in these systems began. 

The initial strategy was to observe all confirmed WASP systems without assigning specific priority, adding newly confirmed planets to the sample as they came in. This strategy was abruptly interrupted in November 2014, when the upgrade of the CORALIE fibers introduced an offset in the systemic velocity zero point. This offset creates a strong degeneracy with the slope of any long-term radial velocity trend, making the detection of new companions significantly more difficult. It was therefore decided to restrict observations to systems in which an additional companion or trend had already been identified. Another offset was introduced in November 2024 following the replacement of the thorium-argon calibration lamp. The three instrumental versions are hereafter referred to as CORALIE07, CORALIE14, and CORALIE24. As of 2026, only 22 systems remain under observation. 

Several companions to H-J identified through this program have already been published. These include the companions to WASP-41\,b and WASP-47\,b \citep{Neveu2016}, WASP-53\,b and WASP-81\,b \citep{Triaud2017}, WASP-134\, \citep{Anderson2018}, and WASP-132\,b \citep{Grieves2025}.

\subsection{Targets}

We retrieved all WASP systems observed with CORALIE from the DACE platform\footnote{\url{https://dace.unige.ch/}} using the \texttt{arvi}\footnote{\url{https://github.com/j-faria/arvi}} Python package, identifying 159 targets matching the pattern ``WASP-*'' in the database. We then cross-matched this list with the NASA Exoplanet Archive\footnote{\url{https://exoplanetarchive.ipac.caltech.edu/}} \citep[NEA; ][]{NEA2025} and the \textit{PlanetS} catalog\footnote{\url{https://dace.unige.ch/exoplanets/}} \citep{Otegi2020,Parc2024} to exclude systems hosting non-planetary objects or false positives, including eclipsing binaries and brown dwarfs, leaving a final sample of 155 targets. This cross-match also allowed us to compile the planetary and stellar parameters needed for our analysis. The systems from \citet{Anderson2014} and \citet{Anderson2018} are absent from both catalogs and were therefore retrieved manually. For planets present only in the NEA, we selected the most appropriate reference and applied the same cross-matching and parameter computation procedure used for the PlanetS catalog, ensuring a homogeneous dataset. Table~\ref{tab:sample} lists the full sample along with the corresponding references and parameters.

\subsection{Radial velocity reduction and post-processing}

CORALIE data were reduced using the standard CORALIE Data Reduction Software (DRS) pipeline (version 3.4 for CORALIE07 and 3.8 for CORALIE14 and CORALIE24), with the CCF mask selected to match the spectral type of each target. Points that did not pass the DRS quality control were discarded. We then applied additional post-processing steps: sigma clipping was performed on the radial velocity measurements to remove outliers (at 5$\sigma$) and on the associated uncertainties to reject the lowest signal-to-noise epochs (at 8$\sigma$). Furthermore, data points taken during transit were removed to avoid contamination by the Rossiter-McLaughlin effect. We removed targets with fewer than 10 radial velocity epochs, as well as instrument versions contributing fewer than 3 measurements, to avoid degeneracies with the instrumental offsets, leaving a final sample of 147 targets. The radial velocity measurements span from June 2007 to November 2025. 

A more complete description of CORALIE, its updates, and DRS can be found in \citep{Figueira2026}.


\section{Homogeneous radial velocity analysis}\label{sect:rv_analysis}

We analyzed the radial velocity datasets using the \texttt{kima}\footnote{\url{https://github.com/kima-org/kima}} package \citep{Faria2018}. This code models the radial velocities as a sum of Keplerians by sampling from the joint posterior distribution using a Diffusive Nested Sampling algorithm \citep[DNS;][]{Brewer2011}, which also provides the Bayesian evidence $\ln\mathcal{Z}$ for model comparison. DNS further supports trans-dimensional sampling \citep{Brewer2015}, in which the number of planets is treated as a free parameter.

\subsection{Priors}

The systemic velocity is modeled using CORALIE07 as the reference instrument, with the offsets of CORALIE14 and CORALIE24 fitted as free parameters. An instrumental jitter term is included for each dataset.

The transiting H-J in each system is incorporated into \texttt{kima} as a known transiting object, with Gaussian priors on the orbital period and time of conjunction taken from the reference paper. We then allow for between 0 and 4 additional planets, with broad uninformative priors on their parameters. The prior for the orbital period is limited to the time span of observations of each target to avoid degeneracies between any long-term trend and the offsets between CORALIE versions. However, if a trend is significantly identified, we increased the limit to allow the search of the possible period/mass of the signal. A Student-$t$ likelihood is adopted to provide a more robust treatment of outliers.

All priors are summarized in Table~\ref{tab:priors}.

\begin{table}[h]
\small
\caption{Prior distribution used in the RV analysis}
\centering
\renewcommand{\arraystretch}{1.3}
\begin{center}
\begin{tabular}{lcc}
\hline\hline
Parameter & Units & Prior \\
\hline
$v_\mathrm{syst, COR07}$ & \ms & $\mathcal{G}(\tilde{RV},\sigma_\mathrm{RV})$ \\
$\mu_\mathrm{j, COR07}$ (offset) & \ms & $\mathcal{G}(0,150)$ \\
$\sigma_{j}$ (jitter) & \ms & $\mathcal{HG}(20)$ \\
\hline
$P_\mathrm{b}$ & days & $\mathcal{G}(P_{HJ},\sigma_{P_{HJ}})$\\
$T_\mathrm{0,b}$ & BJD & $\mathcal{G}(T_{0,HJ},\sigma_{T_{0,HJ}})$\\
$K_b$ & \ms &  $\mathcal{LU}(0.1,2000)$ \\
$e_b$ & & $\mathcal{K}(0.867,3.03)$ \\
$\omega_b$ & rad & $\mathcal{U}(0,2\pi)$ \\
\hline
N$_\mathrm{p}$ & & $\mathcal{U}_{d}$ (0,4) \\
$P_\mathrm{i}$ & days & $\mathcal{U}(10,T_\mathrm{span})$ \\
$K_i$ & \ms &  $\mathcal{LU}(0.1,1000)$ \\
$e_i$ & & $\mathcal{K}(0.867,3.03)$ \\
$\omega_i$ & rad & $\mathcal{U}(0,2\pi)$ \\
$M_{0,i}$ & rad & $\mathcal{U}(0,2\pi)$ \\
\hline
\end{tabular}
\label{tab:priors}
\end{center}
\begin{tablenotes}
    \item \textbf{Notes.} $\tilde{RV}$ is the mean radial velocity of the timeseries, $P_{HJ}$ and $T_{0,HJ}$ is the orbital period and time of conjunction of the H-J Distributions are abbreviated as follows: $\mathcal{G}$ = Gaussian; $\mathcal{HG}$ = half-Gaussian; $\mathcal{U}$ = uniform, subscript $d$ indicates a discrete distribution; $\mathcal{LU}$ = log-uniform; $\mathcal{K}$ = Kumaraswany.
\end{tablenotes}
\end{table}

\subsection{Stellar activity}

One challenge when searching for long-period companions using the radial velocity technique is stellar activity. In particular, stellar magnetic cycles can mimic the signal of a long-period planetary companion \citep[e.g.][]{Lovis2011}, and must be carefully accounted for to avoid false detections. 
When a signal is detected, we inspect the activity indicators available in the CORALIE pipeline, namely the FWHM, BIS, H$_\alpha$, and Na index, to check whether an equivalent periodicity is present. Measurements of the Ca\,\textsc{ii} H\&K emission would provide a more direct tracer of magnetic field strength and spot and plage coverage, but cannot be reliably extracted from CORALIE spectra due to the insufficient signal-to-noise ratio in the blue part of the spectrum, where contamination from the calibration lamp is too strong.
As an alternative tracer of magnetic cycles, we make use of long-term photometry, which has been shown to carry signatures of stellar magnetic activity cycles \citep{Suarez2016, Montet2017}. For targets showing signs of activity, we retrieve light curves from the ASAS-SN survey\footnote{\url{https://asas-sn.osu.edu/}} \citep{Shappee2014,Kochanek2017}, recomputed using aperture photometry  corrected for proper motion to avoid the spurious trends that a fixed aperture can produce for stars with large proper motion. We then search for long-term periodicities by fitting both a Gaussian process with a quasi-periodic kernel and a simple sinusoidal model, using an MCMC sampler with \texttt{emcee}, with independent offsets applied to each ASAS-SN camera.

\subsection{Results}

For each system, we compared models of increasing complexity using the Bayesian evidence $\ln\mathcal{Z}$, starting from a model containing only the H-J, then successively adding a linear trend, a quadratic trend, and the model with up to four additional Keplerian signals. We refer to the model containing the H-J plus one additional Keplerian as the "one-Keplerian" model, and similarly for the "two-Keplerian" model, and so on. A model is considered significant with respect to the simpler alternative when $\Delta$ln$Z >$ 3, and strongly significant when $\Delta$ln$Z >$ 5 \citep{Feroz2011}. In the following, we present the confirmed companions, the detected trends, and the more ambiguous signals separately.

\subsubsection{Detected outer companions}\label{subsubsect:detected_comp}

We here list the systems for which a model including one or more additional companions is strongly favored by the Bayesian evidence, and for which we can confidently confirm the presence of a planetary or sub-stellar object. This includes both detections reported here for the first time, companions previously characterized by this program, and companions announced based on partial orbits, for which our extended CORALIE baseline now provides complete orbital coverage in most cases, significantly improving the precision on their parameters. For each system, we report the median of the marginal posterior distributions, computed from samples of the selected model with the favored number of planets $N_p$. Uncertainties are defined as the distance between the median and the bounds of the 68.27\% Highest Density Interval (HDI). The orbital and physical parameters for all confirmed companions are summarized in Table~\ref{tab:orbital_params}. \\

\noindent\textit{WASP-8:} 

The presence of an additional companion to WASP-8~b was first reported by \citet{Knutson2014}, though only a fraction of the orbit was covered at the time. Our continued CORALIE observations now span slightly more than two full orbits, allowing us to fully characterize the companion. We find an orbital period of 3054.9$^{+12.9}_{-4.8}$ days, a semi-amplitude of 115.77$^{+1.01}_{-1.40}$~\ms, and an eccentricity of 0.147$^{+0.008}_{-0.010}$, corresponding to a minimum mass of 8.33$^{+0.33}_{-0.29}$~\mjup. 

Our analysis marginally favors a two-companion model over the single Keplerian with $\Delta$ln$Z \sim$~3.14. The second signal has a poorly constrained period, with a maximum likelihood value for the orbital period near 2544~days, a semi-amplitude of 15~\ms, and a high eccentricity of 0.63. Since this orbit would cross that of the confirmed companion, we consider it unlikely to represent a real planet. A stellar magnetic cycle remains a possible explanation, though no corresponding periodicity is detected in the activity indicators or the long-term ASAS-SN photometry.\\

\noindent\textit{WASP-22:} 

A long-term trend was already noted in the discovery paper of WASP-22~b \citep{Maxted2010}. Our CORALIE observations now cover more than two full orbits of the outer companion, revealing a period of 2540$^{+37}_{-30}$ days, a semi-amplitude of 77.4$^{+3.2}_{-3.0}$~\ms, and an eccentricity of 0.461$^{+0.028}_{-0.032}$, corresponding to a minimum mass of 4.89$^{+0.98}_{-0.87}$~\mjup. \\

\noindent\textit{WASP-34:} 

A clear long-term trend was reported in the discovery paper of WASP-34~b \citep{Smalley2011}. Our analysis reveals a well-characterized companion with a period of $\sim$2040~days, consistently recovered across all models. Beyond this first companion, our data marginally favor both a two-Keplerian model ($\Delta\ln\mathcal{Z} \sim 3.36$) and a model combining one Keplerian with a quadratic trend ($\Delta\ln\mathcal{Z} \sim 3.23$) over the single Keplerian solution, No counterpart to the second signal is detected in the activity indicators or the long-term ASAS-SN photometry, suggesting a planetary origin is plausible. Interestingly, in the two-Keplerian model the periods of the two companions are close to a 3:2 mean-motion resonance, which would be a remarkable configuration worth investigating with additional data, but the three alternatives are indistinguishable at this stage. The parameters for each scenario are summarized in Table~\ref{tab:wasp34}. \\

\begin{table}
    \centering
    \small
    \caption{Orbital parameters of the outer companion(s) to WASP-34~b under three model assumptions.}
    \label{tab:wasp34}
    \renewcommand{\arraystretch}{1.2}
    \begin{tabular}{lcc}
        \hline\hline
        & Keplerian c & Keplerian d / trend \\
        \hline
        \multicolumn{3}{l}{\textit{1 Keplerian} ($\Delta\ln\mathcal{Z} = 0$, ref.)} \\
        $P$ (days)        & $2033.4^{+3.6}_{-6.5}$    & - \\
        $K$ (\ms)         & $86.3^{+1.7}_{-1.1}$      &  -\\
        $e$               & $0.214^{+0.009}_{-0.014}$  &  -\\
        $M\sin i$ (\mjup) & $5.29^{+0.21}_{-0.28}$    & - \\
        \hline
        \multicolumn{3}{l}{\textit{1 Keplerian + trend} ($\Delta\ln\mathcal{Z} = 3.23$)} \\
        $P$ (days)              & $2038^{+16}_{-13}$         &  -\\
        $K$ (\ms)               & $79.0^{+2.9}_{-2.8}$       &  -\\
        $e$                     & $0.143^{+0.048}_{-0.030}$  & - \\
        $M\sin i$ (\mjup)       & $4.89 \pm 0.28$            &  -\\
        slope (\msyr)           &      -                      & $4.5^{+1.5}_{-1.1}$ \\
        quad. coeff. (\msyryr)  &    -                        & $-0.31^{+0.27}_{-0.10}$ \\
        \hline
        \multicolumn{3}{l}{\textit{2 Keplerians} ($\Delta\ln\mathcal{Z} = 3.36$)\tablefootmark{a}} \\
        $P$ (days)        & $2058 \pm 12$               & $3083^{+293}_{-238}$ \\
        $K$ (\ms)         & $72^{+11}_{-4}$             & $28.0 \pm 7.7$ \\
        $e$               & $0.182^{+0.032}_{-0.020}$  & $0.243^{+0.050}_{-0.243}$ \\
        $M\sin i$ (\mjup) & $4.58^{+0.47}_{-0.52}$     & $1.24^{+0.52}_{-0.55}$ \\
        \hline
    \end{tabular}
    \tablefoot{\tablefoottext{a}{The maximum likelihood solution provides a significantly better fit than the posterior median.}}
\end{table}

\noindent\textit{WASP-41 \& WASP-47:} 

The companions WASP-41~c and WASP-47~c were first published in \citet{Neveu2016}. No additional data were obtained for WASP-41 after that publication, but WASP-47 obtained few more points. WASP-41 is a moderately active star, and the fit of the ASAS-SN photometry reveals a well-constrained periodicity of $1557 \pm 63$ days. A consistent signal is also visible in the FWHM and H$_\alpha$ index of the CORALIE data.

Our updated parameters for WASP-41~c are: a period of 418.0$^{+2.0}_{-2.9}$ days, a semi-amplitude of 95.1$^{+3.5}_{-3.7}$~\ms corresponding to a minimum mass of $3.18 \pm 0.20$~\mjup, and an eccentricity of 0.287$^{+0.030}_{-0.033}$. For WASP-47~c, we find a period of 578.2$^{+8.2}_{-8.5}$ days, a semi-amplitude of 27.0$^{+5.2}_{-4.4}$~\ms corresponding to a minimum mass of 1.10$^{+0.18}_{-0.20}$~\mjup, and a eccentricity consistent with 0 of 0.12$^{+0.06}_{-0.12}$. \\

\noindent\textit{WASP-53:} 

The brown dwarf companion to WASP-53~b was announced in \citet{Triaud2017} based on an incomplete orbit. Our extended CORALIE baseline now covers nearly two full orbits, yielding a well-constrained solution: a period of 3083.7$^{+5.5}_{-6.5}$ days, a semi-amplitude of 477$^{+12}_{-8}$~\ms corresponding to a minimum mass of 16.67$^{+0.79}_{-0.89}$~\mjup, and an high eccentricity of 0.8365$^{+0.0059}_{-0.0058}$. \\

\noindent\textit{WASP-67:} 

WASP-67~b was discovered by \citet{Hellier2012} with no indication of an additional signal in the radial velocities. We report here the detection of an outer companion with a period of 2244$^{+46}_{-50}$ days, a semi-amplitude of 49.6$^{+4.0}_{-4.7}$~\ms corresponding to a minimum mass of 2.88$^{+0.28}_{-0.31}$~\mjup, and an eccentricity of 0.132$^{+0.077}_{-0.074}$. The time series of the radial velocities and the phase folded at the period of the planets are shown in Fig.~\ref{fig:WASP-67_timeseries}. \\

\noindent\textit{WASP-68:} 

The discovery of WASP-68~b was reported by \citet{Delrez2014}, where the radial velocities marginally favored a linear trend. Continued observations have allowed us to detect and characterize a remarkably long-period, highly eccentric brown dwarf companion with a period of 6211$^{+337}_{-252}$days, a semi-amplitude of 245$^{+8}_{-14}$~\ms corresponding to a minimum mass of 15.35$^{+0.32}_{-0.33}$~\mjup, and an eccentricity of 0.804$^{+0.014}_{-0.028}$. \\

\noindent\textit{WASP-81:} 

The brown dwarf companion WASP-81~c was first characterized by \citet{Triaud2017} using a combination of HARPS and CORALIE data. We recover the detection in our CORALIE-only analysis, though the parameters are less precisely constrained given the absence of the HARPS dataset. We find a period of 1306$^{+11}_{-9}$ days, a semi-amplitude of 1190$^{+15}_{-27}$~\ms corresponding to a minimum mass of 55.6$^{+2.0}_{-2.3}$~\mjup, and an eccentricity of 0.5649$^{+0.0090}_{-0.0076}$. \\

\noindent\textit{WASP-99:} 

WASP-99~b was characterized by \citet{Hellier2014} with no mention of an additional signal. We detect a long-period brown dwarf companion whose orbit is not yet complete, though already well constrained. As the baseline does not yet cover a full orbit, the posterior distributions are asymmetric and the maximum likelihood solution provides a better representation of the data than the posterior median. We find a period of 5795$^{+21}_{-239}$ days, a semi-amplitude of 189.6$^{+5.7}_{-1.2}$~\ms corresponding to a minimum mass of 21.4$^{+1.1}_{-1.0}$~\mjup, and an eccentricity of 0.202$^{+0.008}_{-0.014}$. \\

\noindent\textit{WASP-132:} 

As part of this program, \citet{Grieves2025} announced the presence of a cold giant companion to the H-J WASP-132~b and its inner super-Earth WASP-132~c discovered by \citet{Hellier2017} and \citet{Hord2022}, respectively. That paper also reported an additional long-term trend, which we continue to monitor. Our updated orbital parameters for the companion are: a period of 1741$^{+17}_{-23}$ days, a semi-amplitude of 113.2$^{+6.5}_{-8.9}$~\ms corresponding to a minimum mass of $5.63 \pm 0.40$~\mjup, and an eccentricity of 0.171$^{+0.063}_{-0.048}$. The linear trend has a slope of 19.8$^{+1.7}_{-1.4}$~\msyr. \\

\noindent\textit{WASP-134:} 

The peculiar WASP-134 system, hosting a H-J alongside a companion on a relatively short 70-day orbit, was reported by \citet{Anderson2018}. A small number of additional measurements have been obtained since publication. Additional power is present in the periodogram at longer periods, and a model including a linear trend is significantly favored over the one-Keplerian solution with $\Delta\ln\mathcal{Z} \sim 4.3$. We therefore adopt the one-Keplerian plus trend model, which gives a 70.196$^{+0.055}_{-0.066}$ days, a semi-amplitude of 35.0$^{+2.7}_{-2.4}$~\ms corresponding to a minimum mass of 0.768$^{+0.065}_{-0.055}$~\mjup, and an eccentricity consistent with 0 of 0.054$^{+0.030}_{-0.054}$. \\

\noindent\textit{WASP-185:} 

WASP-185 hosts an eccentric H-J discovered by \citet{Hellier2019}. Our analysis strongly favors the inclusion of an additional signal ($\Delta\ln\mathcal{Z} \sim 5.93$), with a period of 1072~days and a semi-amplitude of 65~\ms. Strikingly, when this companion is included in the model, the eccentricity of the H-J becomes consistent with zero ($e = 0.06$), in contrast to the eccentric solution obtained without it ($e = 0.19$). This raises the interesting possibility that the apparent eccentricity of the H-J was previously biased by the unmodeled companion signal. No counterpart is detected in the activity indicators or the ASAS-SN photometry, supporting a planetary interpretation. We caution, however, that our dataset covers only a single orbit of the companion, and further observations are needed to confirm the detection. The parameters for both models are summarized in Table~\ref{tab:wasp185}. \\

\begin{table}
    \centering
    \small
    \caption{Orbital parameters of WASP-185 system under two model assumptions.}
    \label{tab:wasp185}
    \renewcommand{\arraystretch}{1.2}
    \begin{tabular}{lcc}
        \hline\hline
        Parameter & HJ only & HJ + companion \\
        \hline
        \multicolumn{3}{l}{\textit{WASP-185~b}} \\
        $K_\mathrm{b}$ (\ms)         & $86.2^{+9.0}_{-8.3}$   & $63.3^{+5.4}_{-6.8}$ \\
        $e_\mathrm{b}$               & $0.191^{+0.076}_{-0.059}$ & $0.058^{+0.009}_{-0.058}$ \\
        $M_\mathrm{b}$ (\mjup) & $0.94 \pm 0.10$         & $0.702^{+0.073}_{-0.071}$ \\
        \hline
        \multicolumn{3}{l}{\textit{WASP-185~c}} \\
        $P_\mathrm{c}$ (days)        &              -           & $1061^{+61}_{-55}$ \\
        $K_\mathrm{c}$ (\ms)         &            -             & $63^{+7}_{-10}$ \\
        $e_\mathrm{c}$               &              -           & $0.097^{+0.013}_{-0.097}$ \\
        $M_\mathrm{c}\sin i$ (\mjup) &              -           & $3.24^{+0.51}_{-0.43}$ \\
        \hline
    \end{tabular}
\end{table}

\subsubsection{Detected trends}\label{subsubsect:detected_trends}

We list here the systems for which a linear or quadratic radial velocity drift is significantly favored over the baseline H-J model, but where the signal cannot yet be attributed to a fully characterized Keplerian orbit. Such trends are indicative of a massive companion on a long-period orbit, though in some cases a stellar activity origin cannot be excluded. Where possible, we discuss the constraints that can already be placed on the companion's period and mass. \\

\noindent\textit{WASP-5:} 

A weak trend is detected in the radial velocities of WASP-5, whose H-J was discovered by \citet{Anderson2008}. The detection is only moderately significant ($\Delta$ln$Z \sim$~3.12), with a slow drift of $-5.25 \pm 1.39$~\msyr. Given the fitted jitter level of $\sim$10~\ms over more than eight years of observations, it is difficult to firmly distinguish between a very long-period massive companion and a stellar magnetic cycle. No corresponding clear signal is detected in the activity indicators. \\

\noindent\textit{WASP-15:} 

A clear linear trend is visible in the CORALIE07 radial velocities of WASP-15, a system whose H-J was first reported by \citet{West2009}, but it does not persist in the CORALIE14 dataset. Our analysis favors a quadratic trend model, with a slope of 2.39$^{+0.63}_{-0.77}$~\msyr and a quadratic coefficient of $-0.329 \pm 0.082$~\msyryr. Despite over 15 years of observations, the interpretation remains ambiguous: the apparent trend may be degenerate with the offset between the CORALIE07 and CORALIE14 datasets, while the relatively large CORALIE14 jitter of $\sim$8~\ms leaves open the possibility that the signal is of stellar origin. No corresponding variation is detected in the activity indicators. \\

\noindent\textit{WASP-38:} 

WASP-38 \citep{Barros2011} shows a clear quadratic trend, with a slope of 4.1$^{+2.8}_{-1.9}$~\msyr and a quadratic coefficient of 3.70$^{+0.72}_{-0.57}$~\msyryr. Despite the clear detection of the trend, its origin is uncertain. The radial velocity trend has no clear counterpart in the activity indicators. The fitted jitter of $\sim$14~\ms is comparable in amplitude to the trend itself, suggesting that stellar activity could plausibly account for the observed signal. \\

\noindent\textit{WASP-108:} 

WASP-108~b was announced by \citet{Anderson2014}. We detect a clear and significant linear trend of 23.42$^{+0.85}_{-0.76}$~\msyr, with no corresponding variability in the activity indicators or long-term photometry. This signal is consistent with the stellar companion discovered through high-contrast imaging with VLT/SPHERE by \citet{Bohn2020}, who found a star of $0.35 \pm 0.02$~\msol at $\sim$ 0.124'', corresponding to a projected separation of $\sim$ 32~AU. \\

\noindent\textit{WASP-171:} 

A linear trend was already noted in the discovery paper of WASP-171~b \citep{Nielsen2019}. Our extended baseline now reveals a clear quadratic trend, with a slope of -22.3$^{+3.3}_{-3.0}$~\msyr and a quadratic coefficient of $8.6 \pm 1.0$~\msyryr. Given the advanced curvature of the signal, we also searched for a Keplerian solution, finding a best-fit period of 4350$^{+355}_{-585}$ days and a semi amplitude of 168$^{+24}_{-30}$~\ms with eccentricity of 0.34$^{+0.16}_{-0.12}$. However, the posteriors on the period remains highly asymmetric; we find that the minimum period at 3$\sigma$ is 3478 days. Taking the maximum likelihood values, the inferred minimum mass is 12.87$^{+2.28}_{-2.85}$~\mjup, which itself should be regarded as a lower limit given the incomplete orbit. The companion most likely falls above the planetary regime. \\

\subsubsection{Ambiguous signals}\label{subsubsect:ambiguous}

We list here the systems where additional periodic signals are present in the radial velocities, but whose origin remains uncertain. These include cases where the signal properties are consistent with either a long-period companion or a stellar magnetic cycle, as well as marginal detections that do not reach our significance threshold. Some of these signals were identified by inspecting overdensities in the posterior distribution of the mass-period plane relative to the prior, which provides a useful diagnostic for weak or poorly constrained signals that could affect our occurrence rate determination in Sect.~\ref{sect:occurrence_rates}. \\

\noindent\textit{WASP-7:} 

WASP-7 hosts a H-J first reported in \citet{Hellier2009}. The star is a fast rotator with $v\sin i \sim 17$~\kms, reflected in the large CCF FWHM of 26~\kms, as expected for an F5V star, and has an activity level of $\log R'_\mathrm{HK} = -4.76$ estimated from HARPS spectra. We detect a significant signal in the radial velocities near 1203~days with a semi-amplitude of $37 \pm 5$~\ms. Some power is also present in the 1200-1400~day range in the FWHM of the CORALIE data, while the ASAS-SN photometry shows large scatter without a clearly identified period. We cannot exclude that this signal arises from a magnetic cycle, given the fast rotation and the level of activity of this star. \\

\noindent\textit{WASP-44:} 

WASP-44 was discovered by \citet{Anderson2012}. We detect a marginal signal at long periods, close to the edge of our observational baseline and more consistent with a trend than a resolved Keplerian. The ASAS-SN photometry suggests a periodicity near 1155~days that could be associated with this signal, and corresponding signatures are also visible in the activity indicators, making a stellar origin plausible. \\

\noindent\textit{WASP-46:}

WASP-46 was published by \citet{Anderson2012}, who noted emission in the Ca\,\textsc{ii} H\&K lines and a rotational modulation in the WASP photometry, both pointing to an active star. We detect signals near 25~days and in the 800-1500~day range in the radial velocities. The ASAS-SN photometry suggests a possible periodicity near 1200~days broadly consistent with these signals. Given the clear activity signatures in this system, a stellar origin seems favored. \\

\noindent\textit{WASP-79:}

WASP-79 is an F-type star hosting a H-J on a polar orbit \citep{Smalley2012, Addison2013}. It rotates rapidly, with a CCF FWHM of $\sim$28~\kms, though its mean $\log R'_\mathrm{HK} = -4.78$ from HARPS data does not indicate a particularly active star despite the fast rotation. We detect a signal near 200-300~days with an amplitude of $\sim$30~\ms that has no counterpart in the activity indicators. The ASAS-SN photometry shows a bimodal period distribution with peaks near 400 and 750~days. The polar orbit of the H-J could be indicative of past dynamical interactions in the system, which would make a planetary origin for this additional signal plausible. However, given the fast rotation, a short magnetic cycle cannot be excluded. \\

\noindent\textit{WASP-84:} 

WASP-84 hosts a H-J orbiting a G9V/K0V star \citep{Anderson2014b}. The host star shows strong Ca\,\textsc{ii} H\&K emission with $\log R'_\mathrm{HK} = -4.44$, and the WASP photometry reveals a rotational modulation near 14~days. We detect an additional signal near 516~days with an amplitude of $20 \pm 5$~\ms, though it is only marginally significant ($\Delta\ln\mathcal{Z} \sim 2.11$). While this period is somewhat short for a stellar magnetic cycle \citep[e.g.,][]{Chahal2025}, we cannot completely exclude it given the high activity level of the star and the scatter visible in the activity indicators. More observations are needed to clarify the nature of this signal. \\ 

\noindent\textit{WASP-89:} 

WASP-89~b was reported by \citet{Hellier2015}, who noted clear magnetic activity signatures including rotational modulation and star spot crossings during transit. Our analysis favors a one-Keplerian model, but the period is poorly constrained, with multiple peaks near 1700 and 3000~days. The offsets between CORALIE07 and CORALIE14 also appear correlated with these solutions, adding further uncertainty. Some scatter is visible in the activity indicators. The semi-amplitude of $102 \pm 38$~\ms seems large from a magnetic cycle, but our dataset is insufficient to confirm or refute the presence of a real companion.  \\

\noindent\textit{WASP-111:} 

WASP-111 \citep{Anderson2014} is a fast rotator with a CCF FWHM of $\sim$17~\kms and variable activity levels, with a transition from an active to a quiescent state noted between the 2012 and 2013 observing seasons. Our analysis favors a one-Keplerian model, but the period is entirely unconstrained and multiple signals appear to coexist in the dataset, most likely reflecting the high level of stellar activity, with an additional correlation visible between the FWHM and the RVs.  \\

\subsubsection{Refined hot Jupiter parameters}

In addition to the companion search, our extended observational baseline allows us to refine the orbital parameters of the H-Js themselves. However, since we rely solely on CORALIE data, the parameters are not improved for all systems, as many have been characterized with additional instruments offering better precision. We therefore provide updated H-J parameters only for systems where our analysis yields a lower relative mass uncertainty than the reference listed in Table~\ref{tab:sample}, as well as for all systems where a companion or trend was detected, since the presence of an additional signal can influence the derived parameters. Updated parameters for these 64 H-Js are given in Table~\ref{tab:hj_params}, fully available online.
We note that the planetary masses are derived from the stellar masses taken from the reference papers, as refining stellar parameters for the full sample is beyond the scope of this work. The semi-amplitude K is therefore the more robust quantity to consider. We also compared our derived H-J masses across the full sample to the reference values, finding agreement within 1$\sigma$ for all but 7 systems, as shown in Fig.~\ref{fig:mass_comparison}. \\

The seven discrepant H-J masses are those of WASP-26~b, WASP-126~b, WASP-156~b, WASP-162~b, WASP-166~b, WASP-174~b, and WASP-185~b. For WASP-126, WASP-156, WASP-162, and WASP-174, the discrepancy reflects the limits of CORALIE's characterization capacity at the low-mass end: all four planets have masses below 0.35~\mjup, a regime where CORALIE's precision is insufficient and where the reference papers typically rely on HARPS or other higher-precision instruments. For WASP-26~b, we find a slightly lower mass (0.99~\mjup) than the reference value of 1.04~\mjup from \citet{Bonomo2017}, which benefits from additional HARPS-N measurements. For WASP-166~b, we find a mass of $4.55 \pm 0.15$~\mjup, somewhat lower than the $5.20 \pm 0.20$~\mjup reported by \citet{Hellier2019}, possibly related to a marginal non-zero eccentricity found in our solution that is not present in the reference. Finally, for WASP-185~b, due to the detected companion and the updated eccentricity (see Sect.~\ref{subsubsect:detected_comp}), we find a mass of $0.702 \pm 0.072$~\mjup lower than the $0.980 \pm 0.060$~\mjup from \citet{Hellier2019}.\\


\section{Occurrence rates of outer companions}\label{sect:occurrence_rates}

\subsection{Method}

\begin{figure}[t]
\centering
\includegraphics[width=0.40\textwidth]{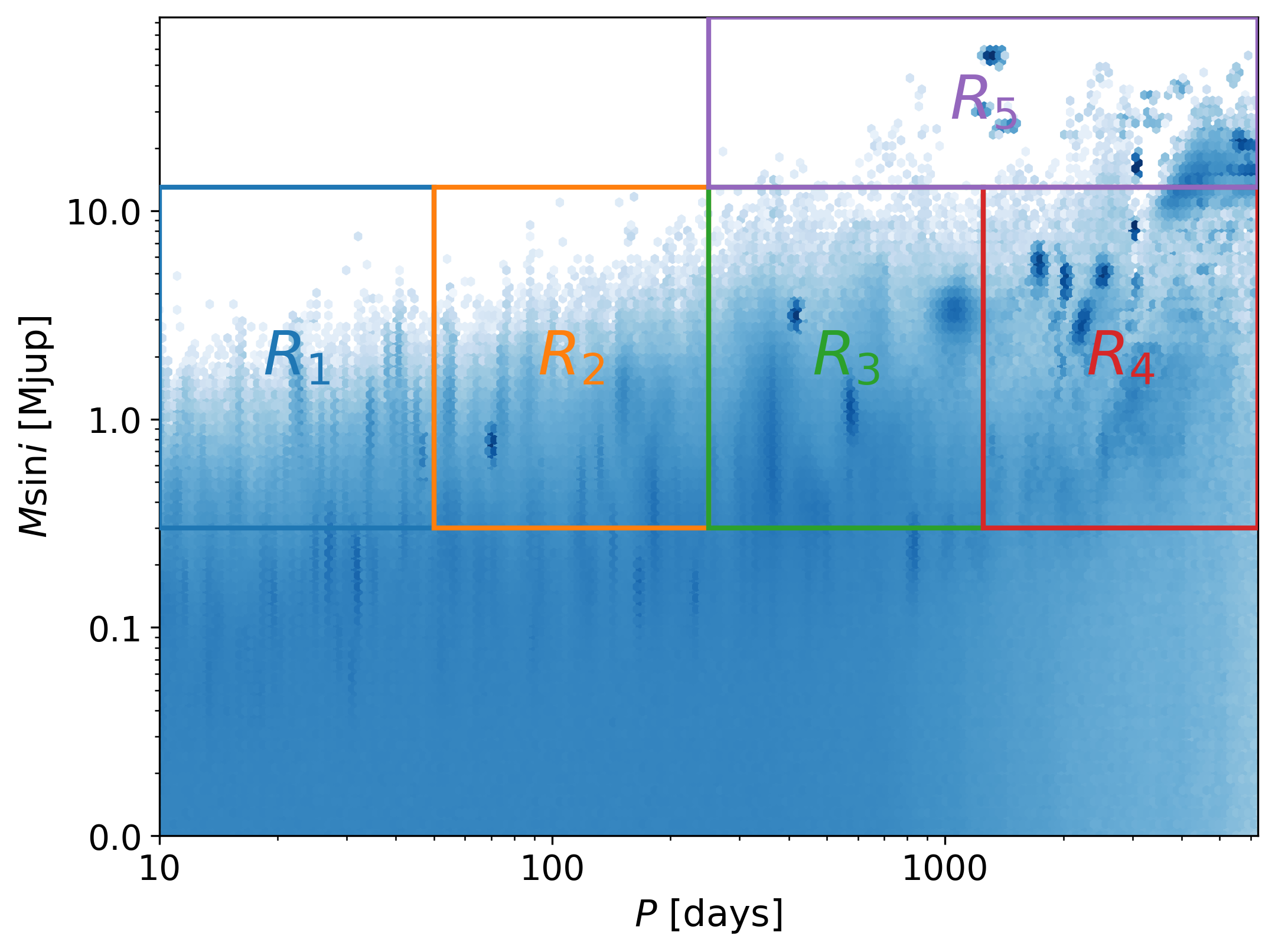}
\caption{Joint posteriors for the minimum mass and orbital period, shown as hexbin density plots, for the restricted sample. All samples with N$_\mathrm{p}$ > 0 are shown. The five regions where the occurrence rates are estimated are shown as colored boxes and labels.}
\label{fig:post_boxes}
\end{figure}

We derive the occurrence rate of outer companions to H-Js using the framework introduced by \citet{Faria2025}\footnote{\url{https://github.com/kima-org/occurrence}}. The method combines posterior samples from the individual \texttt{kima} analyses of each system to infer a posterior distribution for the occurrence rate within a given region of parameter space, defined by bounds in orbital period and minimum mass. Rather than relying on injection-recovery tests to estimate detection limits or requiring an explicit detection threshold, the approach uses importance sampling to re-weight the posterior samples from each system according to the fraction of samples that fall within the region of interest. The occurrence rate $f_R$ is defined as the fraction of stars hosting at least one companion within the region, and its posterior is evaluated on a grid after combining the contributions from all systems in the sample. 

A key advantage of this approach is that systems with no detected companion still contribute information through the upper portion of their posterior distributions, which constrain the mass and period range accessible to the data. The method therefore naturally accounts for the varying sensitivity of each dataset without the need to compute a separate completeness map.

We define five regions of interest in the period-mass plane. Regions $R_1$ to $R_4$ probe the planetary companion regime (0.3--13~\mjup) across four successive period bins equally spaced in logarithm: 10--50~days, 50--250~days, 250--1250~days, and 1250--6250~days. Region $R_5$ covers the brown dwarf regime (13--85~\mjup) over the broader period range of 250--6250~days. The prior probability of each region, $f_0$, defined by the individual priors used in the analysis of each star, is estimated from \texttt{kima} using Monte Carlo sampling, following \citet{Faria2025}.

We first derive the occurrence rates using all 147 systems in our sample, with the exception of WASP-4, whose H-J exhibits orbital decay \citep{Basturk2025} that is not accounted for in our modeling and would bias the results. We then assess the sensitivity of the results to the treatment of ambiguous cases by running the analysis on a restricted sample that excludes systems where a detected drift cannot be confidently attributed to a companion rather than stellar activity, namely WASP-5, WASP-15, and WASP-38 (see Sect.~\ref{subsubsect:detected_trends}), as well as all systems with ambiguous signals discussed in Sect.~\ref{subsubsect:ambiguous}. For WASP-41, where stellar activity is present alongside a confirmed companion, we account for this by including a Gaussian process component in the radial velocity model rather than excluding the system. The joint posteriors for the minimum mass and orbital period for this restricted sample, together with the five regions of interest, are shown in Fig.~\ref{fig:post_boxes}.

\subsection{Results}

\begin{figure*}[!htbp]
\centering
\includegraphics[width=\textwidth]{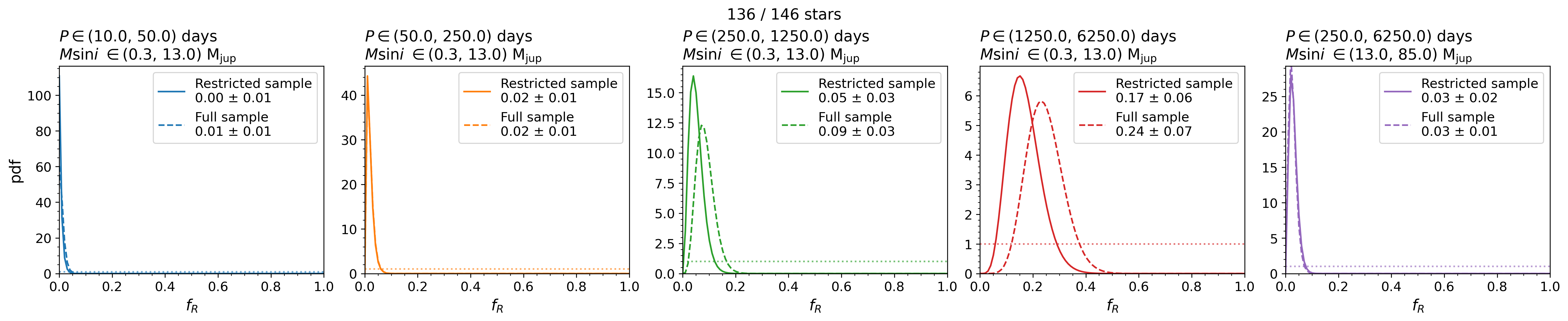}
\caption{Occurrence rate posterior distributions for the five regions $R_1$--$R_5$, for the full (dashed line) and restricted sample (solid line).  Each panel also shows the mean and standard deviation of the posterior distribution for both samples. The prior is shown as a dotted line, it has a mean and a standard deviation of 0.5 and 0.28, respectively.}
\label{fig:occu}
\end{figure*}

The occurrence rate posterior distributions for the five regions are shown in Fig.~\ref{fig:occu} for both the full and restricted samples.

In the short-period bins $R_1$ (10--50~days) and $R_2$ (50--250~days), the occurrence rates are very low, with $f_{R_1} = 0.01 \pm 0.01$ and $f_{R_2} = 0.02 \pm 0.01$ for the full sample, consistent with the restricted sample values ($f_{R_1} < 0.01$ at 1$\sigma$ and $f_{R_2} = 0.02 \pm 0.01$). The single detection driving the $R_2$ result is WASP-134~c, the only confirmed companion in this period range. The posteriors in these two regions are close to zero and well constrained, indicating that close-in massive companions to H-Js are rare in our sample.

The occurrence rate increases significantly at longer periods. In $R_3$ (250--1250~days), we find $f_{R_3} = 0.09 \pm 0.03$ for the full sample and $f_{R_3} = 0.05 \pm 0.03$ for the restricted sample, and in $R_4$ (1250--6250~days), $f_{R_4} = 0.24 \pm 0.07$ and $f_{R_4} = 0.17 \pm 0.06$ respectively. 

Deriving the occurrence for the entire $R_1-R4$ region, we find $f_{R_1-R_4} = 0.14 \pm 0.03$ for the full sample and $f_{R_1-R_4} = 0.10 \pm 0.03$ for the restricted sample.

While the two samples remain consistent within $1\sigma$, the lower values in the restricted sample are expected: the excluded systems host signals in these period ranges that could not be confidently attributed to a real companion. We therefore consider the restricted sample to provide a more conservative but robust estimate.

For the brown dwarf regime $R_5$ (250--6250~days, $13$--$85$~\mjup), we find $f_{R_5} = 0.03 \pm 0.01$ for the full sample and $f_{R_5} = 0.03 \pm 0.02$ for the restricted sample, consistent with each other and indicating a scarcity of brown dwarf companions to H-Js.


\section{Discussion}\label{sect:discussion}

\begin{figure*}[t]
\sidecaption
\includegraphics[width=10cm]{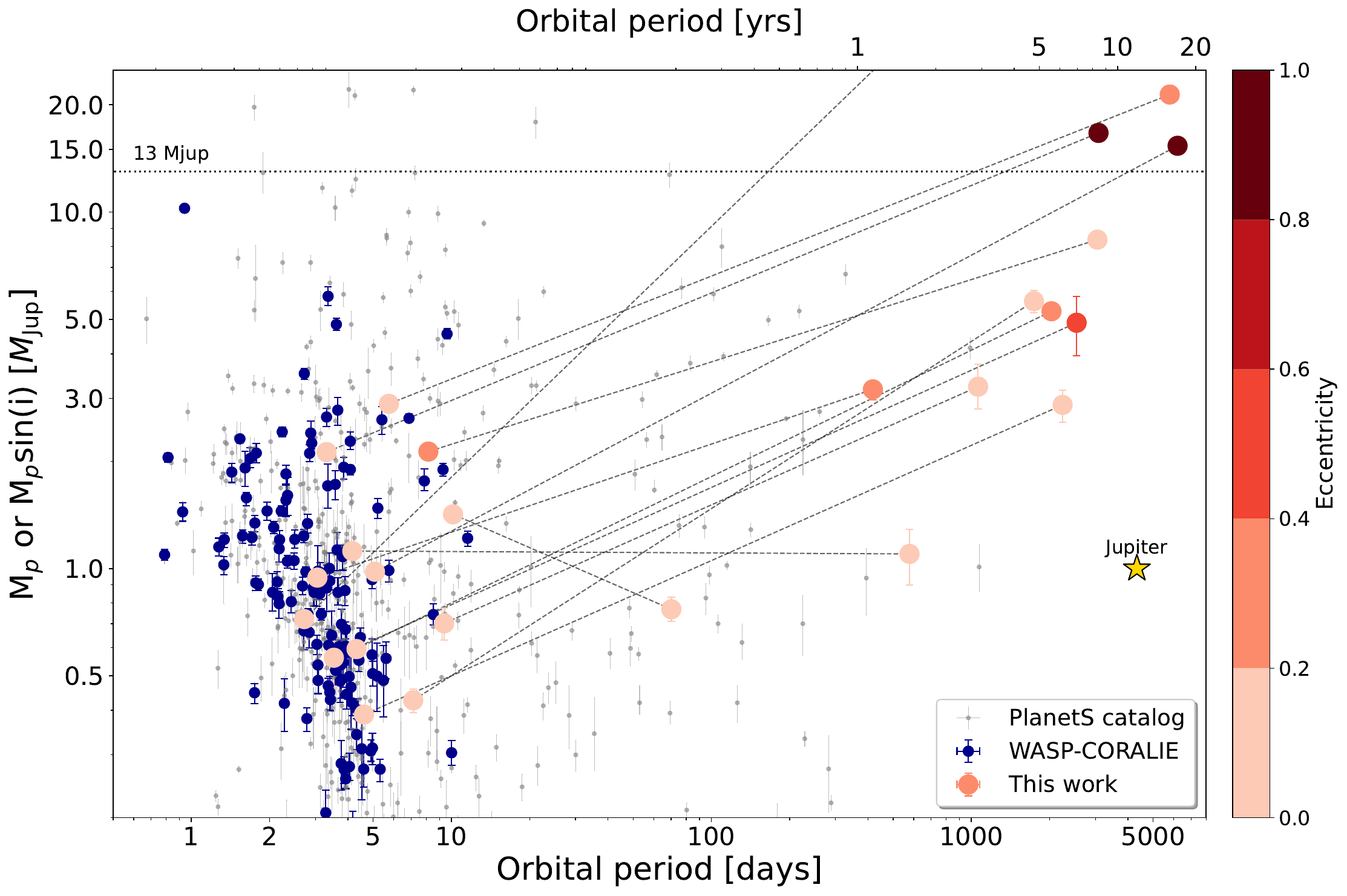}
\caption{Mass or minimum mass versus orbital period diagram. Grey dots show the well-characterized exoplanets from the \textit{PlanetS} catalog \citep{Otegi2020,Parc2024}; blue dots show all the hot Jupiters from our WASP-CORALIE survey. Hot Jupiters and their detected companions from this study are color-coded by eccentricity and linked, system by system, with a dashed grey line.}
\label{fig:mass_period_CORALIE}
\end{figure*}

We report the detection or refinement of 13 outer giant companions to H-Js and derive the associated occurrence rates across orbital period for our WASP--CORALIE survey. Fig.~\ref{fig:mass_period_CORALIE} summarizes the minimum masses and orbital periods of both the H-Js in our survey and their detected companions. In this section, we examine the parameters of the H-Js and their companions and search for correlations between them, test whether H-J and host star properties depend on the presence of a companion, compare our occurrence rates to the literature, and assess the detectability of our detected signals with \textit{Gaia} DR4/DR5.

\subsection{Parameters exploration and correlations}\label{subsect:params_explore}

To investigate possible correlations between the properties of H-Js and those of their outer companions, we performed pairwise Spearman rank correlation analyses across four parameters for both the H-J and its companion: orbital period, semi-major axis, eccentricity, and planetary mass. This gives sixteen parameter combinations, letting us search for links between the orbital architecture and physical properties of the inner and outer planets. We built a combined sample of systems with a H-J and an outer companion by merging the systems identified in this study with those reported in the literature, selecting from the NEA H-Js in multi-planet systems with at least one outer giant companion satisfying $P_c > 20$ days and $M_c\mathrm{sin}i_c > 0.3$~\mjup. This yields 30 H-Js with detected companions, three of which have two outer companions. As in Fig.~\ref{fig:mass_period_CORALIE}, Fig.~\ref{fig:mass_period_all} shows this sample in a mass-period diagram.

Measurement uncertainties were propagated with a Monte Carlo approach: for each system we bootstrapped the measured quantities assuming Gaussian uncertainties over 5,000 realizations, computing the Spearman coefficient ($\rho$) and its p-value each time. We adopted the median $\rho$ as our best estimate and used the fraction of realizations with $p < 0.05$ as a measure of how robust each correlation is.

\begin{figure}[t]
\centering
\includegraphics[width=0.35\textwidth]{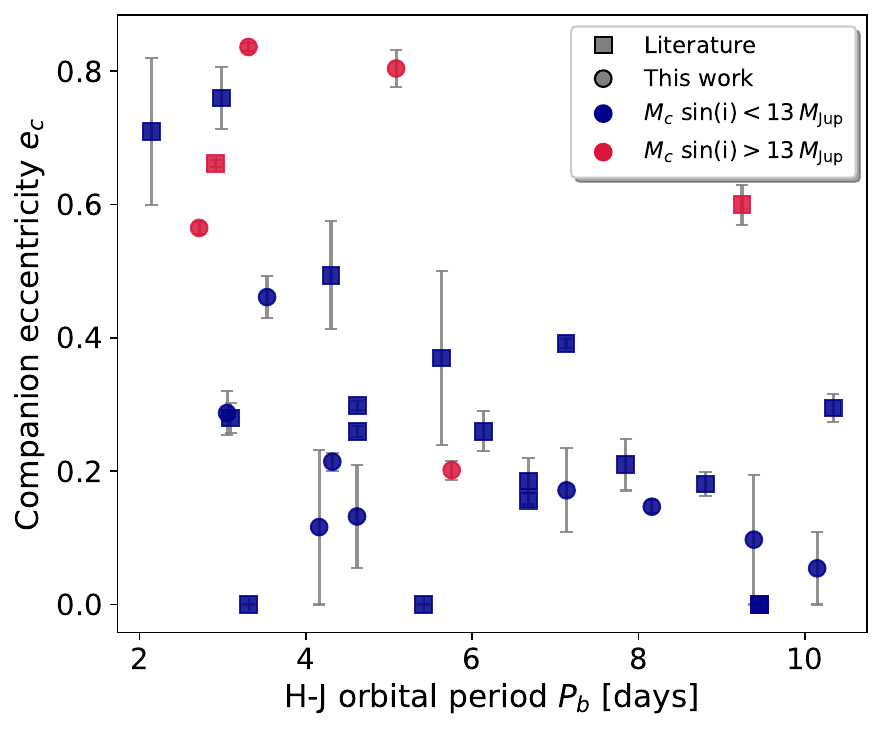}
\caption{Companion eccentricity as a function of the hot Jupiter's orbital period, for the combined sample of 30 companion-hosting systems from this study (circles) and the literature (squares). Companions with minimum masses above 13~\mjup\ is highlighted in red.}
\label{fig:corr_ecc_period}
\end{figure}

Of the sixteen combinations, only one meets our significance criterion of $p < 0.05$ in at least 95\% of realizations: a moderate anti-correlation between the H-J's orbital period and its companion's eccentricity, with $\rho=-0.51^{+0.03}_{-0.04}$. Shorter-period H-Js tend to have companions on more eccentric orbits, while companions of longer-period H-Js are typically closer to circular. Fig.~\ref{fig:corr_ecc_period} shows this relation, with companions with masses above 13~\mjup\ highlighted in red. Visually, the distribution is better described by a triangular envelope than by a simple linear trend: short-period H-Js span the full range of companion eccentricities, from circular to highly eccentric, whereas long-period H-Js are consistently paired with low-eccentricity companions, so that an upper boundary in eccentricity decreases with increasing period. The few points that depart from this envelope correspond to particularly massive and eccentric companions.

This anti-correlation is consistent with high-eccentricity migration (HEM), in which a companion's eccentricity sets the strength of the Kozai--Lidov or the secular forcing it exerts on the H-J, driving it to a smaller pericenter distance; since angular momentum is approximately conserved during the subsequent tidal circularization, this pericenter sets the H-J's final period \citep{Fabrycky2007,Naoz2012,Dawson2018}. Our result, a shorter final period for H-Js with more eccentric companions, matches this expectation and adds to the observational evidence supporting HEM as a formation channel for at least part of the H-J population. We also caution that this links the companion's present-day eccentricity to an excitation event that may have occurred long ago, so the correlation observed today would understate the one imprinted at the time of migration if the companion's eccentricity has since evolved.

\begin{figure*}[t]
\centering
\sidecaption
\includegraphics[width=12cm]{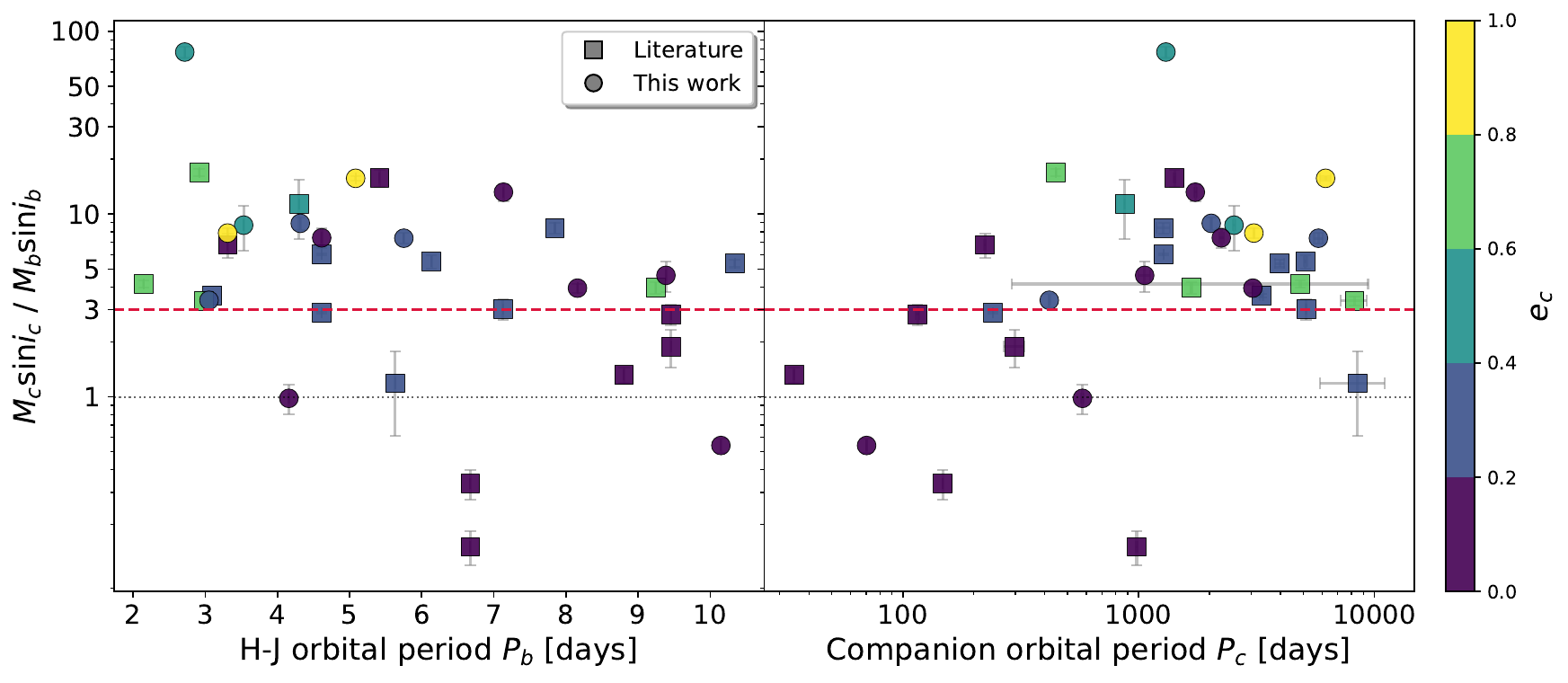}
\caption{Mass ratio $M_c\mathrm{sin}i_c/M_b\mathrm{sin}i_b$ between the outer companion and the hot Jupiter, as a function of the hot Jupiter's orbital period (left) and the companion's orbital period (right), color-coded by companion eccentricity. The horizontal red dashed line marks the mass ratio of 3 from the \citet{Zink2023} criterion, and the horizontal grey dotted line marks a mass ratio of 1.}
\label{fig:mass_ratio_period}
\end{figure*}

\citet{Zink2023} recently found that an eccentricity above 0.2 and a mass of the outer companion a factor 3 higher are key ingredients in H-J formation, both markers of a companion capable of driving strong dynamical forcing on the inner planet, consistent with the high-eccentricity migration picture discussed above. In both Fig.~\ref{fig:mass_period_CORALIE} and Fig.~\ref{fig:mass_period_all}, the companion is almost always more massive than its H-J. Fig.~\ref{fig:mass_ratio_period} plots the mass ratio $M_c\mathrm{sin}i_c/M_b\mathrm{sin}i_b$ against the orbital periods of both planets, color-coded by companion eccentricity.

Two populations stand out here. Hot Jupiters with periods below 5 days have systematically higher mass ratios, with 11/15 systems meeting the Zink \& Howard criterion, against 7/18 above 5 days. The same split appears when using the companion's period instead: below 1000 days mass ratios mostly stay under 3 (3/12 meet the criterion), while above 1000 days almost every system exceeds it (15/21). Since the H-Js in this sample cover a fairly narrow range of period and mass, this points to the more massive companions sitting preferentially at wider separations. This is confirmed directly by the correlation between mass ratio and period ratio $P_c/P_b$: $\rho = 0.39^{+0.03}_{-0.02}$ for the combined literature sample, rising to $\rho = 0.65^{+0.04}_{-0.03}$ for our survey alone (Fig.~\ref{fig:corr_mass_period_ratio}). WASP-134 and HIP\,14810 \citep{Ment2018} are the only two systems with a mass ratio below one; HIP\,14810 is especially unusual, with two outer companions both less massive than its H-J. 

We caution that we used minimum masses for both the outer companions and the H-Js entering this mass ratio. Only six H-Js are non-transiting, so their true mass is well constrained; the outer companions, by contrast, are mostly unconstrained in inclination (only three benefit from an astrometric constraint from Hipparcos-Gaia), so an unknown, possibly large inclination could shift individual systems across the threshold once true masses are known. Detection bias at longer periods also favors the discovery of larger-mass companions, so smaller-mass companions may still be present in HJ systems but remain undetected.

As explained in \citet{Zink2023}, this combination of a mass ratio above 3 and a moderately eccentric outer companion, seen preferentially among our wider-separation systems, points towards a specific HEM mechanism: coplanar high-eccentricity migration \citep{Petrovich2015}. Compared to Kozai--Lidov cycles operating at high mutual inclination, it requires a near-coplanar geometry. This mechanism specifically predicts a mass ratio above 3 and a residual companion eccentricity of 0.2--0.5, matching what we find. Mutual inclinations, where available, would help disentangle the two migration sub-channels; astrometric orbits from \textit{Gaia} DR4 and DR5 (see Sect.~\ref{subsect:detec_gaia}) should provide such constraints for a fraction of our companions in future work.

\subsection{Dependence of system properties on companion presence}

While Sect.~\ref{subsect:params_explore} examined correlations between H-J and companion parameters within companion-hosting systems, we now ask whether H-J properties themselves differ between systems with and without a detected companion. We therefore compared the properties of H-Js hosting at least one detected outer giant companion with those of the overall H-J population. We used the same companion-hosting sample as the previous sub-section and retrieved from the NEA the global H-J population with $0.3~\mjup < M_b\sin(i) < 13~\mjup$ and $P < 11$ days. After removing the companion-hosting systems from this parent sample, the reference population consists of 586 H-J systems without identified companions. We compared both planetary and stellar parameters between the two samples: orbital period, eccentricity, and mass of the H-Js, together with the effective temperature, radius, mass, metallicity, and age of the host stars.

We used the Mann--Whitney U test to compare the distributions, and complemented the resulting p-values with Cliff's delta ($\delta$) as a measure of effect size \citep{Cliff1993}. $\delta$ quantifies the degree of stochastic dominance between two groups and ranges from $-1$ to $+1$, with values close to zero indicating largely overlapping distributions and larger absolute values indicating a clearer separation between the two populations. We interpret statistical significance together with effect size throughout, in order to distinguish trends that are physically meaningful from differences driven mainly by the imbalance in sample size.

The clearest difference between the two populations is seen in orbital eccentricity. Hot Jupiters with companions have systematically higher eccentricities than the comparison sample, with median values of 0.025 and 0.0, respectively. The Mann--Whitney test gives a highly significant result ($p = 1.1\times10^{-8}$), with a large effect size ($\delta \simeq 0.50$). This difference is driven mainly by a larger fraction of companion-hosting systems with non-zero eccentricities, rather than a uniform shift across the whole population.

A similarly significant difference appears in orbital period. Companion-hosting H-Js tend to orbit farther from their star than the general population, with median periods of about 5.3 days compared to 3.4 days ($p = 1.4\times10^{-5}$, $\delta \simeq 0.47$, a moderate-to-large effect). Both this shift and the higher eccentricities seen above are naturally explained if the general H-J population, without a detected outer companion, predominantly migrated through the disk-driven channel, which halts at a comparatively fixed inner disk edge and delivers planets to short, near-circular orbits \citep{Lin1996}. Companion-hosting H-Js, by contrast, are more likely to have migrated under the influence of their outer companion, and high-eccentricity migration halts once tidal circularization catches up with the shrinking pericenter set by the companion's eccentricity, a less deterministic stopping point that can leave the H-J at somewhat longer, less fully circularized orbits.

Among the stellar properties, only effective temperature shows a difference between the two populations ($p = 0.019$, $\delta \simeq -0.25$): host stars with both a H-J and a companion have a median temperature of 5611~K, compared to 5900~K for hosts with a H-J alone, though the effect size remains small. No significant differences are found for stellar mass, radius, metallicity, age, or planetary mass. 

\begin{figure*}[t]
\centering
\includegraphics[width=\textwidth]{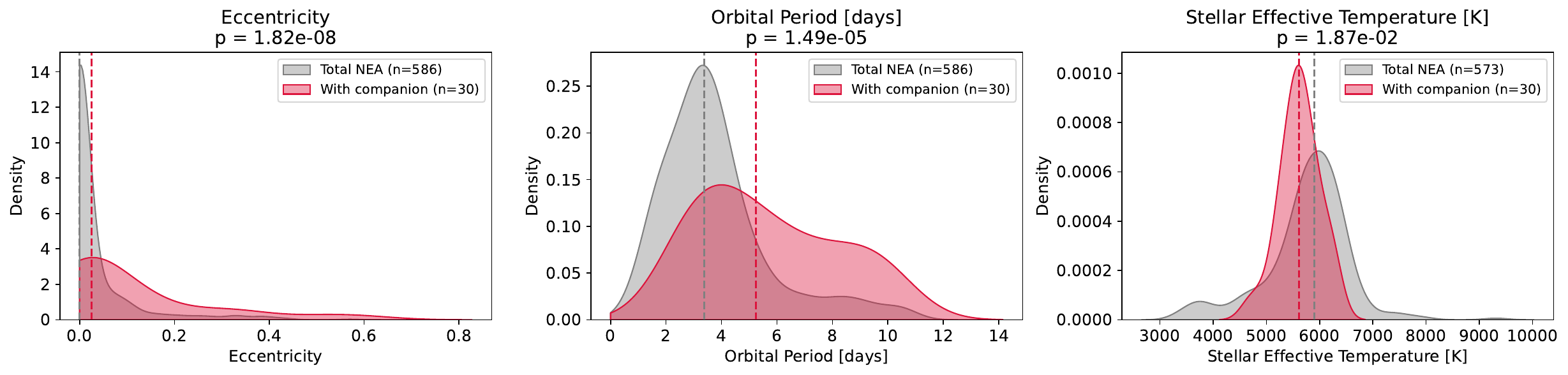}
\caption{Kernel density distributions of the three parameters showing a statistically significant difference between H-Js with a detected companion (red) and those without (grey). Dashed vertical lines indicate the median of each distribution.}
\label{fig:kde}
\end{figure*}

Figure~\ref{fig:kde} shows the distributions of the three parameters found to be statistically significant.

Since the stellar parameters in this comparison come from heterogeneous sources, we repeated it using the homogeneous SWEET-Cat catalog\footnote{\url{https://sweetcat.iastro.pt/}} \citep{Sousa2024}.
Homogeneous stellar parameters from SWEET-Cat were available for 26 companion-hosting host stars and 254 hot-Jupiter-only host stars. With this subsample, the temperature difference becomes stronger ($p = 0.004$, $\delta \simeq -0.34$, a moderate effect size), with median temperatures of 5673~K for the companion sample and 6051~K for the hot-Jupiter-only sample. As in the heterogeneous sample, no significant metallicity difference is found in the homogeneous subsample ($p = 0.19$, $\delta \simeq 0.16$). However, this temperature trend could partly reflect a detection bias rather than a physical difference: cooler stars are typically quieter in RV, making it easier to detect and characterize the low-amplitude, long-period signal of an outer companion, while hotter, earlier-type stars often show stronger magnetic activity cycles that can produce RV signals with periods similar to those of giant companions, complicating their detection and characterization, as illustrated in Sect.~\ref{subsubsect:ambiguous}.

Overall, these results show that companion-hosting H-Js are set apart primarily by their orbital architecture, higher eccentricities and longer periods, rather than by the bulk properties of their host stars or their own mass. This is consistent with the two migration pathways discussed above: H-Js without a detected companion are more likely to have migrated through the disk-driven channel, while companion-hosting systems show the imprint of high-eccentricity migration excited by their outer companion.

\subsection{Occurrence rates in context}

Previous surveys have searched for massive companions to H-Js. Combining long-term RV follow-up with AO imaging, \citet{Knutson2014} reported that massive outer companions are common around H-Js, with an occurrence rate of $51\pm10\%$ for companions with masses between 1 and 13~\mjup\ and orbital separations between 1 and 20~AU. Extending this approach to wider separations, \citet{Bryan2016} combined RV and imaging to probe companions out to $\sim$100~AU, reporting an occurrence rate of $52\pm5\%$ for giant planets between 1 and 20~\mjup\ within 5--20~AU. Both studies relied primarily on long-term radial velocity accelerations, with many companions remaining only partially characterized, and therefore provide an occurrence integrated over a broad region of parameter space rather than as a function of orbital period. Their sensitivity also extends to substantially larger separations than those explored by the present survey, making a direct comparison difficult: for a solar-mass host, our period bins correspond to projected separations of roughly 0.09--0.27 ($R_1$), 0.27--0.78 ($R_2$), 0.78--2.3 ($R_3$), and 2.3--6.6~AU ($R_4$). Where our range overlaps theirs, in $R_3$ and $R_4$, we find occurrence rates well below their integrated values, though consistent with the rising trend we observe with orbital period. One point of agreement with both studies is the near-absence of detected companions within 1~AU, with the exception of WASP-134~c in our sample.

More recently, \citet{Zink2023} used a population-level analysis of the California Legacy Survey to derive occurrence rates for giant exoplanets. They found that once completeness corrections are applied, H-Js host on average $1.3^{+1.0}_{-0.6}$ outer giant companions when integrated over 10--40,000~days. Integrating their companion rate (see their Fig.~3) over our narrower period range yields $0.45^{+0.45}_{-0.28}$, larger than our cumulative occurrence across $R_1$--$R_4$ ($0.14\pm0.03$ for the full sample, $0.10\pm0.03$ for the restricted sample). However, their estimate is based solely on 5 detected companions among their 11 H-J hosts, which explains the large uncertainty. Although slightly below their predicted curve, our occurrence rates are consistent with the slope of their power law, following the same increase in occurrence with orbital period.

\citet{Schlaufman2016} took a different approach, combining previously published RV giant planet statistics with a sample of 136 known RV systems in a Bayesian framework to estimate the probability that a H-J hosts an additional companion, rather than deriving an occurrence rate directly from a homogeneous survey as we do. For companions with $a<5$~AU and $M>1$~\mjup, close to our $R_4$ upper edge, they find a companion probability knowing the presence of an H-J of $0.155^{+0.110}_{-0.077}$, consistent with our cumulative occurrence over $R_1$--$R_4$. 

Finally, it is instructive to compare our results with the occurrence of giant planets around solar-type stars from general radial velocity surveys not selected for the presence of a H-J. \citet{Mayor2011} reported that $9.7\pm1.3\%$ of solar-type stars host a companion more massive than 100~$M_\oplus$ ($\sim 0.3$~\mjup) with orbital periods shorter than 10 years, based on the combined HARPS and CORALIE blind-search samples. \citet{Cumming2008} found a comparable occurrence with $10.5\%$ of FGK stars hosting a planet between 0.3 and 10~\mjup\ over their 2--2000 day survey baseline. Extrapolating their fitted period distribution beyond this baseline, they estimate $11\pm1.7\%$ of FGK stars host a companion between 0.3 and 15~\mjup\ within 5~AU (11~yr), a mass and separation range close to our own. More recently, \citet{Fernandes2019} characterized the orbital period distribution of giant planets ($0.1$--$20~\mjup$) around FGK stars using a broken power law, finding a break near the snow line at $\sim2$--$3$~AU, beyond which the occurrence rate decreases with increasing separation. Integrating their fitted broken power law (see their Table~1) over our period range of 10--6250~days yields an occurrence of $\sim7.6$--$8.2\%$, depending on whether a symmetric or asymmetric break is adopted. This is consistent with their extrapolated occurrence rates for a broader, but overlapping, mass and separation range: $\sim7.9$--$8.9\%$ for $0.1$--$13~\mjup$ companions within 20~AU for the broken-power-law models.

Our cumulative occurrence over $R_1$--$R_4$ are in good agreement with the $\sim8$--$11\%$ occurrence reported for the general FGK population over comparable period and mass ranges. If anything, the full sample lies marginally above this range, which could plausibly reflect the moderately enhanced metallicity expected for our sample: since our stars already host a H-J, and giant planet occurrence is well known to increase with host star metallicity \citep[e.g.,][]{Santos2001, Fischer2005}, a mild enrichment relative to the blind-search field samples underlying \citet{Mayor2011} and \citet{Cumming2008} is expected and would naturally push our occurrence slightly higher. 

Taken together, this suggests that hosting a H-J does not appear to markedly enhance the presence of additional giant companions relative to the wider population of FGK and cold-giant-hosting stars.

\subsection{Detectability with Gaia DR4/DR5}\label{subsect:detec_gaia}

To evaluate whether these companions could in principle be detected astrometrically with \textit{Gaia}, we use the method described by \citet{Lammers2026} described in detail in Appendix~\ref{appendix:gaia}. We apply this analysis both to the detected companions and to the ambiguous signals with a clear periodicity identified in our sample (WASP-7, WASP-79, WASP-84, WASP-89), as astrometric confirmation (or non-detection) could help discriminate between a genuine companion and a stellar activity origin for the latter.

The results are summarized in Table~\ref{tab:gaia_detectability_confirmed}.

Overall, 7 of the 13 confirmed companions (WASP-8, -34, -53, -68, -81, -99, and -132) satisfy the SNR$_1$ detection threshold for both \textit{Gaia} DR4 and DR5, with SNR$_1$ ranging from 2.2 to 11.5. These are predominantly the more massive companions ($M_p\sin i \gtrsim 5~M_J$) at moderate orbital periods, for which the astrometric signal is large enough to be recovered despite only partial orbital phase coverage in some cases (e.g., 32--35\% for WASP-68 and WASP-99 in DR4). The remaining six systems (WASP-22, -41, -47, -67, -134, and -185) fall below the detection threshold in both releases, primarily due to lower planetary minimum masses and, in the case of WASP-134, a very short period that limits the projected semi-major axis. Moreover, with the exception of WASP-8 ($d$ = 90 pc), all these stars are located at distances greater than 130 pc, with a mean distance of 210 pc. In addition, we could expect the detection of the companion responsible for the quadratic trend in WASP-171's RVs if we consider the best-fit model even with partial coverage of the orbit. 

None of the four ambiguous signals with clear periodicity are detectable in DR4. WASP-89 (SNR$_1 = 1.09$) lies between the DR4 and DR5 thresholds and becomes formally detectable only in DR5, making it the most promising candidate among the ambiguous signals for future astrometric confirmation of a planetary origin rather than stellar activity. The other three (WASP-7, WASP-79, and WASP-84) remain undetectable in both releases given their current SNR$_1$ values, so \textit{Gaia} astrometry alone is unlikely to resolve their nature. We note that, for these ambiguous signals, the period and $M_p\sin i$ adopted here correspond to the maximum-likelihood RV solution; given the weak constraints on these parameters, the resulting SNR$_1$ values should be regarded as indicative rather than definitive, and WASP-89 in particular could shift above or below the DR5 threshold depending on the true possible underlying period and mass.

For WASP-34, we treat the astrometric signal of each companion independently under both the single- and two-Keplerian scenarios, rather than jointly modeling their combined astrometric signature; the values reported here therefore correspond to each planet considered in isolation. Under the single-companion model, WASP-34~c reaches SNR$_1 = 2.23$ and is detectable in both DR4 and DR5. Under the two-Keplerian model, the same period is instead attributed to a lower-mass planet (WASP-34~c, SNR$_1 = 1.95$), which remains individually detectable, while the additional companion (WASP-34~d, SNR$_1 = 0.69$) falls below the detection threshold in both releases. \textit{Gaia} astrometry combined with the RVs could help distinguish between the two proposed scenarios.

More generally, for the systems with SNR$_1$ close to the detection thresholds (e.g., WASP-22, -34, -67, -89), we note that combining the astrometric constraint with the existing RVs should substantially improve the effective sensitivity \citep{Baycroft2026}. Throughout this analysis, we have assumed a coplanar configuration between the H-J and its outer companion; a misalignment would increase the true companion mass relative to the minimum mass derived from RVs alone, which could in turn bring some of these near-threshold candidates above the detection limit. Overall, this combination will allow us to derive constraints on the mutual inclination between the planetary and sub-stellar orbits in these systems and refine the orbital parameters of our detected companions.


\section{Conclusions}\label{sect:conclusions}

We have presented a homogeneous long-term radial-velocity analysis of 147 WASP H-J systems observed with CORALIE between 2007 and 2026. By combining individual radial-velocity posteriors with a Bayesian occurrence-rate framework, we have characterized the frequency and properties of additional massive companions while accounting for the heterogeneous sensitivity of the datasets. Our main conclusions are:

\begin{enumerate}

\item We detect or refine 13 outer giant companions to H-Js and measure occurrence rates of companions to WASP H-Js for minimum masses between $0.3$ and $13~\mjup$ over four period bins spanning 10--6250 days. The occurrence rate increases strongly with orbital period, from $<0.01$ and $0.02\pm0.01$ at periods below 250 days to $0.05\pm0.03$ and $0.17\pm0.06$ at longer periods (250--6250 days) for the restricted sample, after removing ambiguous candidates. This shows that additional giant companions to H-Js are preferentially found on multi-year orbits.

\item Over the full explored parameter space, we obtain a cumulative occurrence of $0.14\pm0.03$ for the full sample and $0.10\pm0.03$ for the restricted sample. These values are consistent both with the occurrence of cold giant planets around the general FGK population and, close to previous estimates of outer companions to H-Js, indicating that hosting a H-J does not markedly enhance the presence of additional giant companions.

\item Hot Jupiters hosting outer companions differ from isolated H-Js: they have, on average, longer orbital periods and higher eccentricities. We also find that the companion eccentricity decreases with increasing H-J period, and that the companion-to-hot-Jupiter mass ratio increases with period ratio. These dynamical signatures are consistent with HEM shaping this subset of the H-J population.

\item Finally, \textit{Gaia} astrometry will provide an important complement to this RV monitoring. Seven of the 13 confirmed companions are expected to be detectable with \textit{Gaia} DR4 and DR5, and  will help refine the orbital and planetary parameters and provide constraints on the mutual inclinations of the exoplanets in these systems.

\end{enumerate}

Overall, this large and homogeneous sample confirms that outer giant companions to H-Js cluster preferentially at multi-year orbital separations rather than close-in. When such a companion is present, the H-J itself carries the dynamical imprint of this configuration, with longer orbital periods and higher eccentricities than isolated H-Js, while the companion's own eccentricity decreases with increasing H-J period and the companion-to-hot-Jupiter mass ratio increases with period ratio. Taken together, these findings paint a picture in which HEM shapes a meaningful fraction of the H-J population. With this work, our long-term CORALIE RV monitoring of these systems comes to an end. Combining this dataset with \textit{Gaia} astrometry will nonetheless remain highly valuable for constraining the true masses and inclinations of these companions, and for assessing the dynamical histories of their host systems.

\begin{acknowledgements}
We warmly thank all the observers who contributed to the observations obtained with the Euler/Swiss telescope, whose dedication and careful work made this study possible.
This work has been carried out within the framework of the National Centre of Competence in Research PlanetS supported by the Swiss National Science Foundation under grant 51NF40\_205606. A.De. acknowledges the financial support of the SNSF.\\
PF acknowledges financial support from the Severo Ochoa grant CEX2021-001131-S funded by MCIN/AEI/10.13039/501100011033. PF is also funded by the European Union (ERC, THIRSTEE, 101164189). Views and opinions expressed are however those of the author(s) only and do not necessarily reflect those of the European Union or the European Research Council. Neither the European Union nor the granting authority can be held responsible for them. \\
\end{acknowledgements}

\bibliographystyle{aa} 
\bibliography{bib}

\begin{appendix}

\section{Targets}

\begin{strip}
\centering
\captionof{table}{Stellar and observational parameters of the 147 targets in the CORALIE WASP survey.}
\label{tab:sample}
\renewcommand{\arraystretch}{1.2}
\small
\begin{tabular}{llcccccccccc}
\hline\hline
Host & Reference & $T_{\rm eff}$ & $M_\star$ & $[\rm{Fe/H}]$ & $V_\textrm{mag}$ & $\sigma_{\rm RV}$ & $T_{\rm span}$ & $N_{07}$ & $N_{14}$ & $N_{24}$ & FWHM \\
 & & [K] & [$M_\odot$] & & & [\ms] & [yr] & & & & [\ms] \\
\hline
WASP\mbox{-}2 & Southworth 2010 & $5150^{+80}_{-80}$ & $0.80^{+0.06}_{-0.06}$ & $-0.08^{+0.08}_{-0.08}$ & $11.73$ & $19.8$ & $16.9$ & 42 & 19 & 4 & $7489$ \\
WASP\mbox{-}4 & Baluev et al. 2020 & $5500^{+100}_{-100}$ & $0.93^{+0.05}_{-0.05}$ & \ldots & $12.48$ & $35.5$ & $18.0$ & 41 & 27 & -- & $8386$ \\
WASP\mbox{-}5 & Triaud et al. 2010 & $5700^{+100}_{-100}$ & $1.00^{+0.06}_{-0.06}$ & $0.09^{+0.09}_{-0.09}$ & $12.15$ & $21.5$ & $8.3$ & 38 & 14 & -- & $9035$ \\
WASP\mbox{-}6 & McGruder et al. 2023 & $5438^{+50}_{-50}$ & $0.85^{+0.03}_{-0.02}$ & $-0.15^{+0.05}_{-0.05}$ & $11.91$ & $16.7$ & $7.0$ & 52 & -- & -- & $8228$ \\
WASP\mbox{-}7 & Southworth 2011 & $6400^{+100}_{-100}$ & $1.28^{+0.07}_{-0.07}$ & $0.00^{+0.10}_{-0.10}$ & $9.50$ & $23.9$ & $10.1$ & 87 & 22 & -- & $26433$ \\
\multicolumn{12}{c}{$\vdots$} \\
\multicolumn{12}{c}{\textit{This table is available in its entirety online.}} \\
\hline\hline
\end{tabular}

\vspace{4pt}
\parbox{\textwidth}{\footnotesize
\textbf{Notes.} $T_{\rm eff}$, $M_\star$, and $[\rm{Fe/H}]$ are taken from the discovery reference. $\sigma_{\rm RV}$ is the median RV uncertainty in \ms. $N_{07}$, $N_{14}$, $N_{24}$ are the number of CORALIE measurements with the CORALIE07, CORALIE14, and CORALIE24 versions respectively. The full table is available online.}
\vspace{6pt}
\end{strip}

\section{Systems with companions}

\begin{figure}[h]
    \centering
    \includegraphics[width=0.48\textwidth]{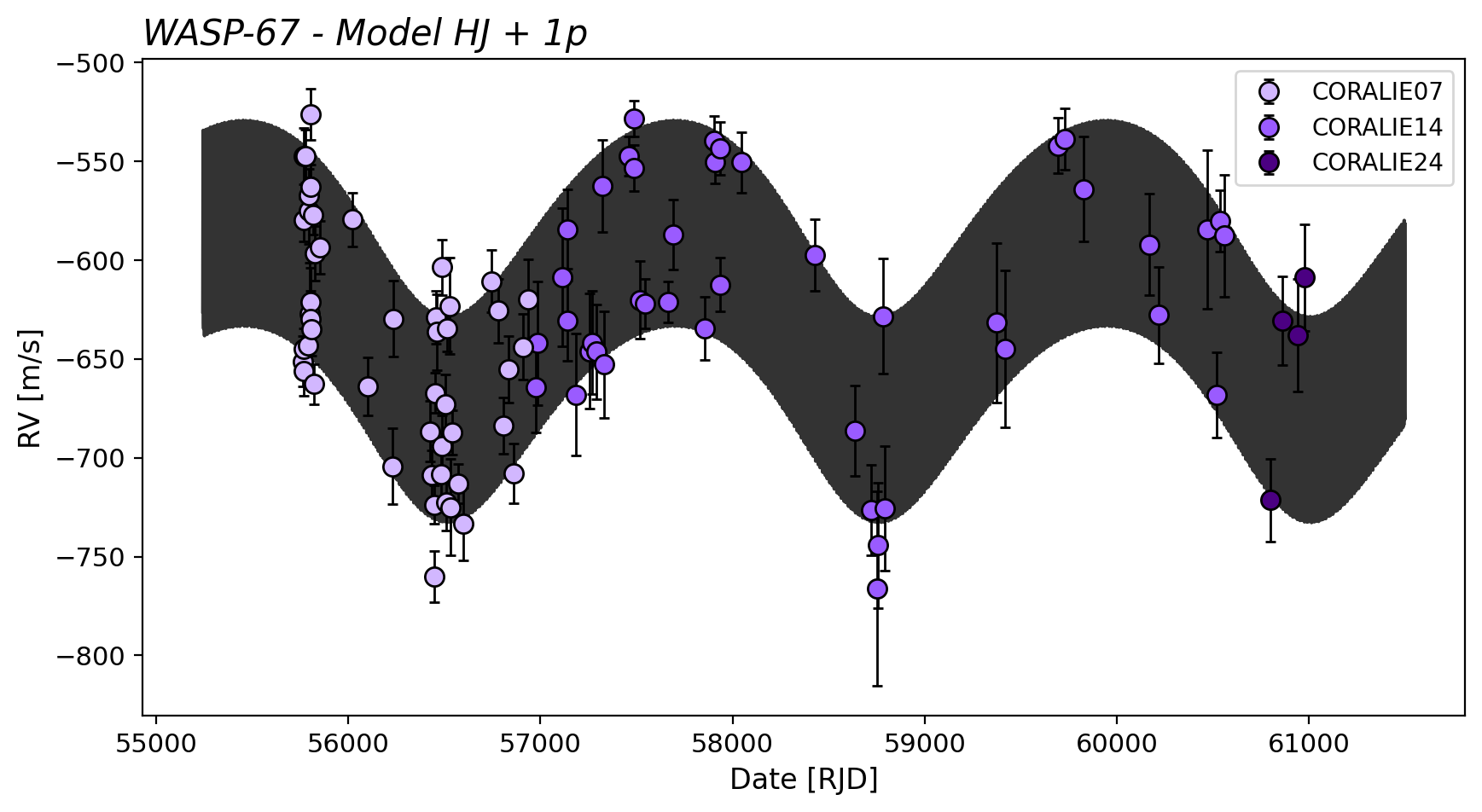}
    \includegraphics[width=0.48\textwidth]{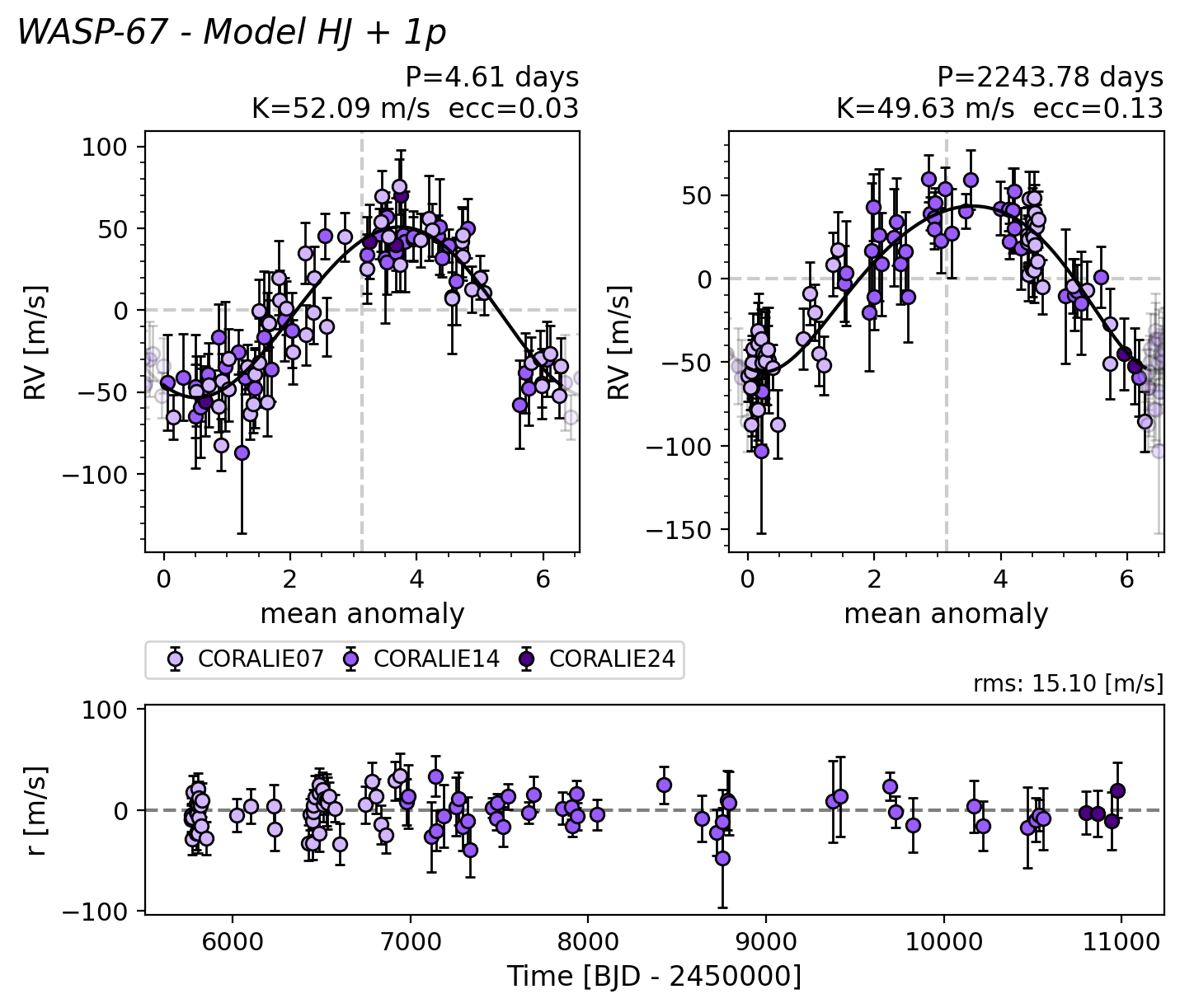}
    \caption{Left: Time series of the CORALIE radial velocities of \mbox{WASP-67} with the best-fit model. Right: Phase folded CORALIE radial velocities of two planets in the WASP-67 system (top) and residuals of the model (bottom). Similar figures for all systems are available online.}
    \label{fig:WASP-67_timeseries}
\end{figure}

\begin{table*}[h]
\centering
\caption{Orbital and physical parameters of the hot Jupiter (planet b) and outer companion (planet c) for each system with confirmed companion.}
\label{tab:orbital_params}
\renewcommand{\arraystretch}{1.2}
\resizebox{\textwidth}{!}{%
\begin{tabular}{lcccccccc}
\hline\hline
Parameter & \textbf{WASP-8} & \textbf{WASP-22} & \textbf{WASP-34} & \textbf{WASP-41} & \textbf{WASP-47} & \textbf{WASP-53} & \textbf{WASP-67} \\
\hline
\multicolumn{8}{l}{\textit{Planet b (hot Jupiter)}} \\
\hline
$P_b$ [d] & $8.158713^{+0.000013}_{-0.000010}$ & $3.532686^{+0.000030}_{-0.000030}$ & $4.3176770^{+0.0000037}_{-0.0000030}$ & $3.0524040^{+0.0000010}_{-0.0000010}$ & $4.159147^{+0.000012}_{-0.000014}$ & $3.3098441^{+0.0000019}_{-0.0000021}$ & $4.6144108^{+0.0000026}_{-0.0000028}$ \\
$T_{0,b}$ [RJD] & $54679.33484^{+0.00033}_{-0.00053}$ & $54780.2502^{+0.0046}_{-0.0039}$ & $54647.55439^{+0.00052}_{-0.00059}$ & $56014.99360^{+0.00010}_{-0.00010}$ & $56982.97720^{+0.00010}_{-0.00010}$ & $55943.56695^{+0.00011}_{-0.00011}$ & $55824.37496^{+0.00023}_{-0.00021}$ \\
$K_b$ [m\,s$^{-1}$] & $222.7^{+1.0}_{-1.3}$ & $70.7^{+2.1}_{-2.2}$ & $73.31^{+0.83}_{-0.89}$ & $138.6^{+2.7}_{-2.9}$ & $140.6^{+2.4}_{-2.6}$ & $324.7^{+3.4}_{-3.1}$ & $52.1^{+2.8}_{-2.4}$ \\
$e_b$ & $0.3132^{+0.0060}_{-0.0017}$ & $0.037^{+0.015}_{-0.037}$ & $0.0048^{+0.0045}_{-0.0048}$ & $0.0090^{+0.0054}_{-0.0090}$ & $0.0070^{+0.0043}_{-0.0070}$ & $0.0101^{+0.0057}_{-0.0064}$ & $0.033^{+0.015}_{-0.033}$ \\
$\omega_b$ [rad] & $4.7864^{+0.0051}_{-0.0028}$ & $1.0^{+0.4}_{-1.0}$ & $3.3^{+2.2}_{-1.4}$ & $2.3^{+2.0}_{-1.5}$ & $2.4^{+0.7}_{-3.1}$ & $3.70^{+0.79}_{-0.68}$ & $2.5^{+1.2}_{-1.1}$ \\
$i_b$ [$^\circ$] & $88.55 \pm 0.16$ & $89.20 \pm 0.50$ & $85.20 \pm 0.20$ & $89.40 \pm 0.30$ & $89.00 \pm 0.17$ & $87.08 \pm 0.15$ & $85.80 \pm 0.35$ \\
$M_b$ [$M_{\rm Jup}$] & $2.147^{+0.075}_{-0.085}$ & $0.56^{+0.11}_{-0.10}$ & $0.592^{+0.030}_{-0.026}$ & $0.942^{+0.053}_{-0.051}$ & $1.119^{+0.043}_{-0.041}$ & $2.123^{+0.083}_{-0.088}$ & $0.389^{+0.023}_{-0.022}$ \\
$a_b$ [AU] & $0.0802^{+0.0020}_{-0.0011}$ & $0.0468^{+0.0048}_{-0.0037}$ & $0.0520^{+0.0009}_{-0.0014}$ & $0.0402^{+0.0010}_{-0.0010}$ & $0.05078^{+0.00085}_{-0.00081}$ & $0.04102^{+0.00082}_{-0.00080}$ & $0.05178^{+0.00077}_{-0.00080}$ \\
\hline
\multicolumn{8}{l}{\textit{Planet c (outer companion)}} \\
\hline
$P_c$ [d] & $3054^{+13}_{-5}$ & $2541^{+37}_{-30}$ & $2033.4^{+3.6}_{-6.5}$ & $418.0^{+2.1}_{-2.9}$ & $578.2^{+8.2}_{-8.5}$ & $3083.7^{+5.5}_{-6.5}$ & $2244^{+46}_{-50}$ \\
$M_{0,c}$ [rad] & $1.429^{+0.055}_{-0.084}$ & $2.450^{+0.069}_{-0.074}$ & $4.490^{+0.059}_{-0.036}$ & $1.15^{+0.13}_{-0.13}$ & $1.2^{+0.5}_{-1.6}$ & $5.528^{+0.011}_{-0.009}$ & $5.43^{+0.64}_{-0.42}$ \\
$K_c$ [m\,s$^{-1}$] & $115.8^{+1.0}_{-1.4}$ & $77.4^{+3.2}_{-3.0}$ & $86.3^{+1.7}_{-1.1}$ & $95.1^{+3.5}_{-3.7}$ & $27.0^{+5.2}_{-4.4}$ & $477^{+12}_{-8}$ & $49.6^{+4.0}_{-4.7}$ \\
$e_c$ & $0.147^{+0.008}_{-0.010}$ & $0.461^{+0.028}_{-0.032}$ & $0.214^{+0.009}_{-0.014}$ & $0.287^{+0.030}_{-0.033}$ & $0.12^{+0.06}_{-0.12}$ & $0.8365^{+0.0059}_{-0.0057}$ & $0.132^{+0.077}_{-0.074}$ \\
$\omega_c$ [rad] & $2.18^{+0.11}_{-0.05}$ & $1.677^{+0.096}_{-0.085}$ & $5.717^{+0.065}_{-0.041}$ & $6.09^{+0.18}_{-0.10}$ & $2.4^{+0.9}_{-1.2}$ & $3.512^{+0.021}_{-0.028}$ & $2.82^{+0.52}_{-0.52}$ \\
$M_c\sin i_c$ [$M_{\rm Jup}$] & $8.33^{+0.33}_{-0.29}$ & $4.90^{+0.99}_{-0.86}$ & $5.29^{+0.21}_{-0.28}$ & $3.18^{+0.20}_{-0.20}$ & $1.10^{+0.18}_{-0.20}$ & $16.67^{+0.79}_{-0.89}$ & $2.88^{+0.29}_{-0.31}$ \\
$a_c$ [AU] & $4.161^{+0.082}_{-0.063}$ & $3.76^{+0.40}_{-0.29}$ & $3.158^{+0.049}_{-0.082}$ & $1.068^{+0.026}_{-0.029}$ & $1.363^{+0.025}_{-0.028}$ & $3.910^{+0.080}_{-0.073}$ & $3.201^{+0.066}_{-0.071}$ \\
\hline
\end{tabular}%
}
\end{table*}

\begin{table*}[h]
\ContinuedFloat
\centering
\caption[]{\textit{(Continued.)}}
\renewcommand{\arraystretch}{1.2}
\label{tab:orbital_params_cont}
\resizebox{\textwidth}{!}{%
\begin{tabular}{lccccccc}
\hline\hline
Parameter & \textbf{WASP-68} & \textbf{WASP-81} & \textbf{WASP-99} & \textbf{WASP-132} & \textbf{WASP-134} & \textbf{WASP-185} \\
\hline
\multicolumn{7}{l}{\textit{Planet b (hot Jupiter)}} \\
\hline
$P_b$ [d] & $5.084300^{+0.000016}_{-0.000012}$ & $2.7164761^{+0.0000023}_{-0.0000023}$ & $5.752612^{+0.000006}_{-0.000039}$ & $7.1335164^{+0.0000018}_{-0.0000020}$ & $10.1467586^{+0.0000078}_{-0.0000084}$ & $9.3851^{+0.0025}_{-0.0024}$ \\
$T_{0,b}$ [RJD] & $56064.86359^{+0.00058}_{-0.00058}$ & $56195.57461^{+0.00020}_{-0.00022}$ & $56224.9841^{+0.0017}_{-0.0017}$ & $59337.60808^{+0.00015}_{-0.00015}$ & $57201.03101^{+0.00072}_{-0.00073}$ & $56936.23^{+0.24}_{-0.25}$ \\
$K_b$ [m\,s$^{-1}$] & $100.2^{+1.2}_{-1.3}$ & $100.1^{+4.4}_{-4.2}$ & $253.9^{+0.86}_{-1.9}$ & $52.9^{+3.7}_{-3.7}$ & $123.5^{+2.5}_{-2.0}$ & $63.3^{+5.4}_{-6.8}$ \\
$e_b$ & $0.012^{+0.004}_{-0.012}$ & $0.049^{+0.017}_{-0.049}$ & $0.0053^{+0.0036}_{-0.0053}$ & $0.051^{+0.020}_{-0.051}$ & $0.113^{+0.018}_{-0.020}$ & $0.058^{+0.009}_{-0.058}$ \\
$\omega_b$ [rad] & $1.5^{+0.2}_{-2.2}$ & $2.50^{+0.71}_{-0.76}$ & $5.0^{+1.2}_{-0.2}$ & $5.0^{+1.5}_{-0.3}$ & $4.622^{+0.051}_{-0.064}$ & $1.5^{+1.0}_{-2.3}$ \\
$i_b$ [$^\circ$] & $88.1 \pm 1.3$ & $88.69 \pm 0.90$ & $88.8 \pm 1.1$ & $89.46 \pm 0.12$ & $89.13 \pm 0.26$ & $86.8 \pm 0.3$ \\
$M_b$ [$M_{\rm Jup}$] & $0.979^{+0.019}_{-0.021}$ & $0.721^{+0.039}_{-0.045}$ & $2.90^{+0.14}_{-0.12}$ & $0.426^{+0.031}_{-0.035}$ & $1.420^{+0.043}_{-0.049}$ & $0.702^{+0.073}_{-0.071}$ \\
$a_b$ [AU] & $0.06217^{+0.00047}_{-0.00053}$ & $0.03909^{+0.00073}_{-0.00072}$ & $0.0716^{+0.0017}_{-0.0015}$ & $0.0670^{+0.0011}_{-0.0012}$ & $0.0956^{+0.0013}_{-0.0012}$ & $0.0905^{+0.0016}_{-0.0016}$ \\
\hline
\multicolumn{7}{l}{\textit{Planet c (outer companion)}} \\
\hline
$P_c$ [d] & $6212^{+337}_{-252}$ & $1306^{+11}_{-9}$ & $5795^{+21}_{-239}$ & $1741^{+17}_{-23}$ & $70.196^{+0.055}_{-0.066}$ & $1061^{+61}_{-55}$ \\
$M_{0,c}$ [rad] & $6.096^{+0.014}_{-0.017}$ & $0.269^{+0.010}_{-0.017}$ & $1.84^{+0.15}_{-0.13}$ & $5.93^{+0.40}_{-0.11}$ & $3.7^{+1.4}_{-1.3}$ & $6.0^{+2.7}_{-1.1}$ \\
$K_c$ [m\,s$^{-1}$] & $245^{+8}_{-14}$ & $1190^{+15}_{-27}$ & $189.6^{+5.7}_{-1.2}$ & $113.2^{+6.5}_{-8.9}$ & $35.0^{+2.7}_{-2.4}$ & $62^{+7}_{-10}$ \\
$e_c$ & $0.804^{+0.015}_{-0.028}$ & $0.5649^{+0.0090}_{-0.0076}$ & $0.202^{+0.008}_{-0.014}$ & $0.171^{+0.063}_{-0.048}$ & $0.054^{+0.030}_{-0.054}$ & $0.097^{+0.013}_{-0.097}$ \\
$\omega_c$ [rad] & $1.992^{+0.028}_{-0.021}$ & $5.595^{+0.015}_{-0.009}$ & $0.15^{+0.09}_{-0.15}$ & $2.28^{+0.22}_{-0.31}$ & $2.8^{+1.3}_{-1.4}$ & $1.0^{+1.0}_{-2.7}$ \\
$M_c\sin i_c$ [$M_{\rm Jup}$] & $15.35^{+0.32}_{-0.34}$ & $55.6^{+2.1}_{-2.3}$ & $21.4^{+1.0}_{-1.0}$ & $5.63^{+0.40}_{-0.40}$ & $0.768^{+0.065}_{-0.055}$ & $3.24^{+0.51}_{-0.43}$ \\
$a_c$ [AU] & $7.11^{+0.25}_{-0.21}$ & $2.400^{+0.046}_{-0.048}$ & $7.14^{+0.20}_{-0.19}$ & $2.619^{+0.051}_{-0.046}$ & $0.3470^{+0.0050}_{-0.0043}$ & $2.122^{+0.086}_{-0.092}$ \\
\hline
\end{tabular}%
}
\end{table*}

\section{Refined hot Jupiter parameters}

\begin{strip}
\centering
\captionof{table}{Updated orbital and physical parameters for the hot Jupiters in the CORALIE WASP survey with improved characterization from our extended baseline. The full table is available online.}
\label{tab:hj_params}
\renewcommand{\arraystretch}{1.2}
\resizebox{\textwidth}{!}{%
\begin{tabular}{lcccccccc}
\hline\hline
Planet & $P_b$ [d] & $T_{0,b}$ [RJD] & $K_b$ [m\,s$^{-1}$] & $e_b$ & $\omega_b$ [rad] & $i_b$ [$^\circ$] & $M_b$ [$M_{\rm Jup}$] & $a_b$ [AU] \\
\hline
WASP\mbox{-}5~b & $1.6284269^{+0.0000039}_{-0.0000032}$ & $54373.99601^{+0.00014}_{-0.00014}$ & $272.2^{+3.2}_{-4.6}$ & $0.012^{+0.006}_{-0.012}$ & $5.0^{+1.7}_{-0.3}$ & $86.1 \pm 1.1$ & $1.579^{+0.073}_{-0.062}$ & $0.02709^{+0.00062}_{-0.00045}$ \\
WASP\mbox{-}7~b & $4.9546415^{+0.0000036}_{-0.0000034}$ & $55446.63493^{+0.00032}_{-0.00029}$ & $94.3^{+3.6}_{-3.5}$ & $0.047^{+0.021}_{-0.047}$ & $4.55^{+0.44}_{-0.40}$ & $87.03 \pm 0.93$ & $0.932^{+0.048}_{-0.050}$ & $0.0618^{+0.0012}_{-0.0011}$ \\
WASP\mbox{-}8~b & $8.158713^{+0.000013}_{-0.000010}$ & $54679.33484^{+0.00033}_{-0.00053}$ & $222.7^{+1.0}_{-1.3}$ & $0.3132^{+0.0060}_{-0.0017}$ & $4.7864^{+0.0051}_{-0.0028}$ & $88.55 \pm 0.16$ & $2.134^{+0.085}_{-0.060}$ & $0.0800^{+0.0019}_{-0.0009}$ \\
WASP\mbox{-}16~b & $3.1186068^{+0.0000011}_{-0.0000013}$ & $54584.42899^{+0.00039}_{-0.00035}$ & $119.2^{+2.4}_{-2.4}$ & $0.013^{+0.006}_{-0.013}$ & $4.9^{+1.9}_{-0.3}$ & $83.99 \pm 0.26$ & $0.850^{+0.034}_{-0.035}$ & $0.04149^{+0.00069}_{-0.00071}$ \\
WASP\mbox{-}18~b & $0.94145245^{+0.00000019}_{-0.00000031}$ & $54664.90530^{+0.00017}_{-0.00016}$ & $1816.5^{+2.6}_{-3.3}$ & $0.0081^{+0.0015}_{-0.0014}$ & $4.74^{+0.10}_{-0.13}$ & $80.6 \pm 1.2$ & $10.25^{+0.23}_{-0.21}$ & $0.02020^{+0.00021}_{-0.00022}$ \\
\multicolumn{9}{c}{$\vdots$} \\
\multicolumn{9}{c}{\textit{This table is available in its entirety online.}} \\
\hline\hline
\end{tabular}%
}
\vspace{6pt}
\end{strip}

\section{Supplementary materials}

\begin{figure*}[!t]
\centering
\begin{minipage}[t]{0.48\textwidth}
  \centering
  \includegraphics[width=0.83\linewidth]{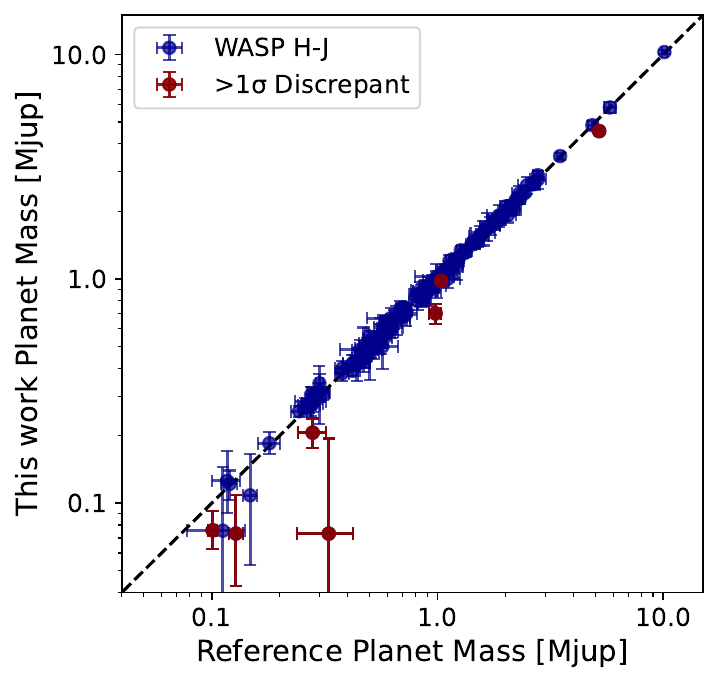}
  \captionof{figure}{Comparison of the minimum masses of the hot Jupiters derived in this work with the published reference values. Red dots indicate measurements discrepant at more than 1$\sigma$ from the reference value.}
  \label{fig:mass_comparison}
\end{minipage}\hfill
\begin{minipage}[t]{0.48\textwidth}
  \centering
  \includegraphics[width=0.83\linewidth]{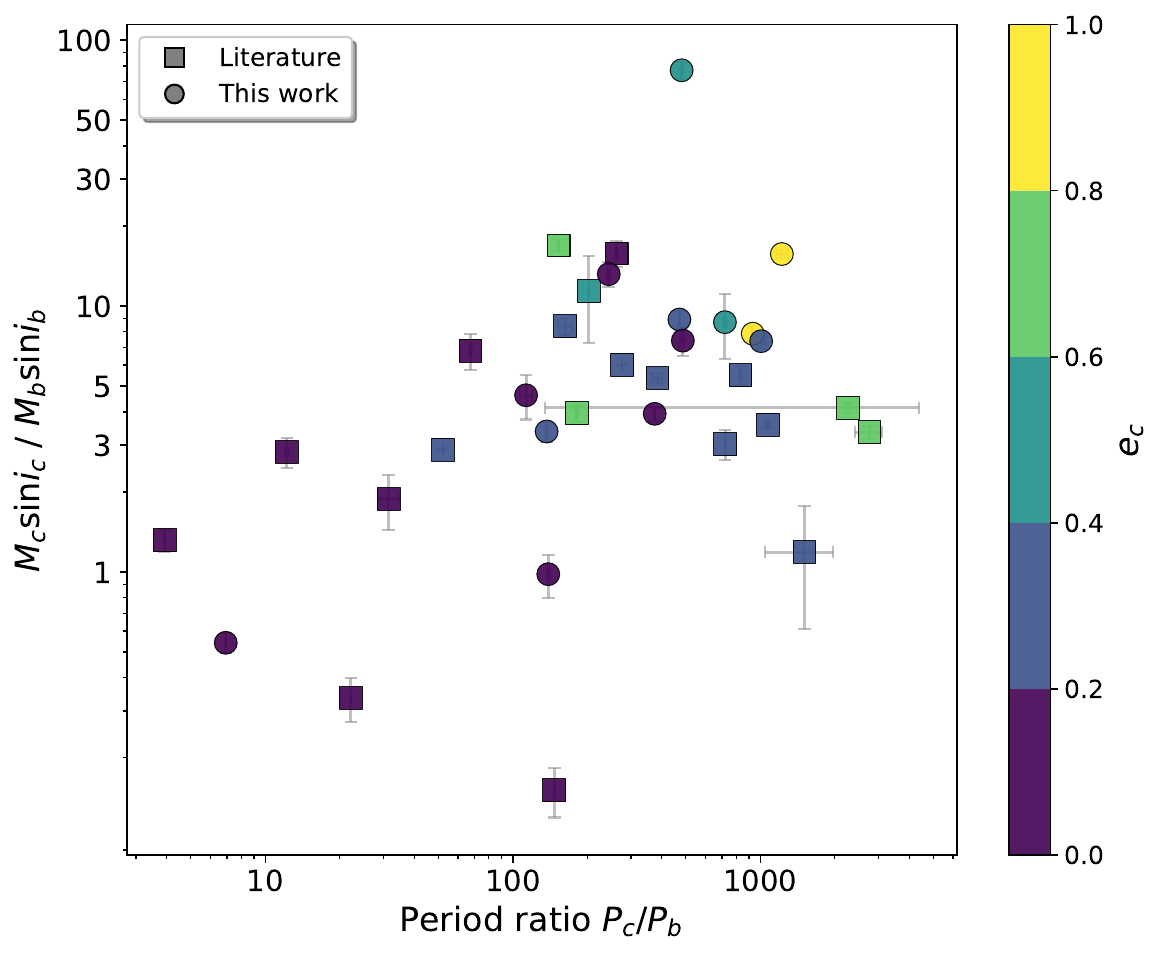}
  \captionof{figure}{Mass ratio $M_c\mathrm{sin}i_c/M_b\mathrm{sin}i_b$ as a function of the period ratio $P_c/P_b$ for the combined sample of 30 companion-hosting systems, color-coded by companion eccentricity. Circles show systems from this work, squares show systems from the literature.}
  \label{fig:corr_mass_period_ratio}
\end{minipage}

\vspace{1em}

\begin{minipage}[c]{12cm}
  \includegraphics[width=12cm]{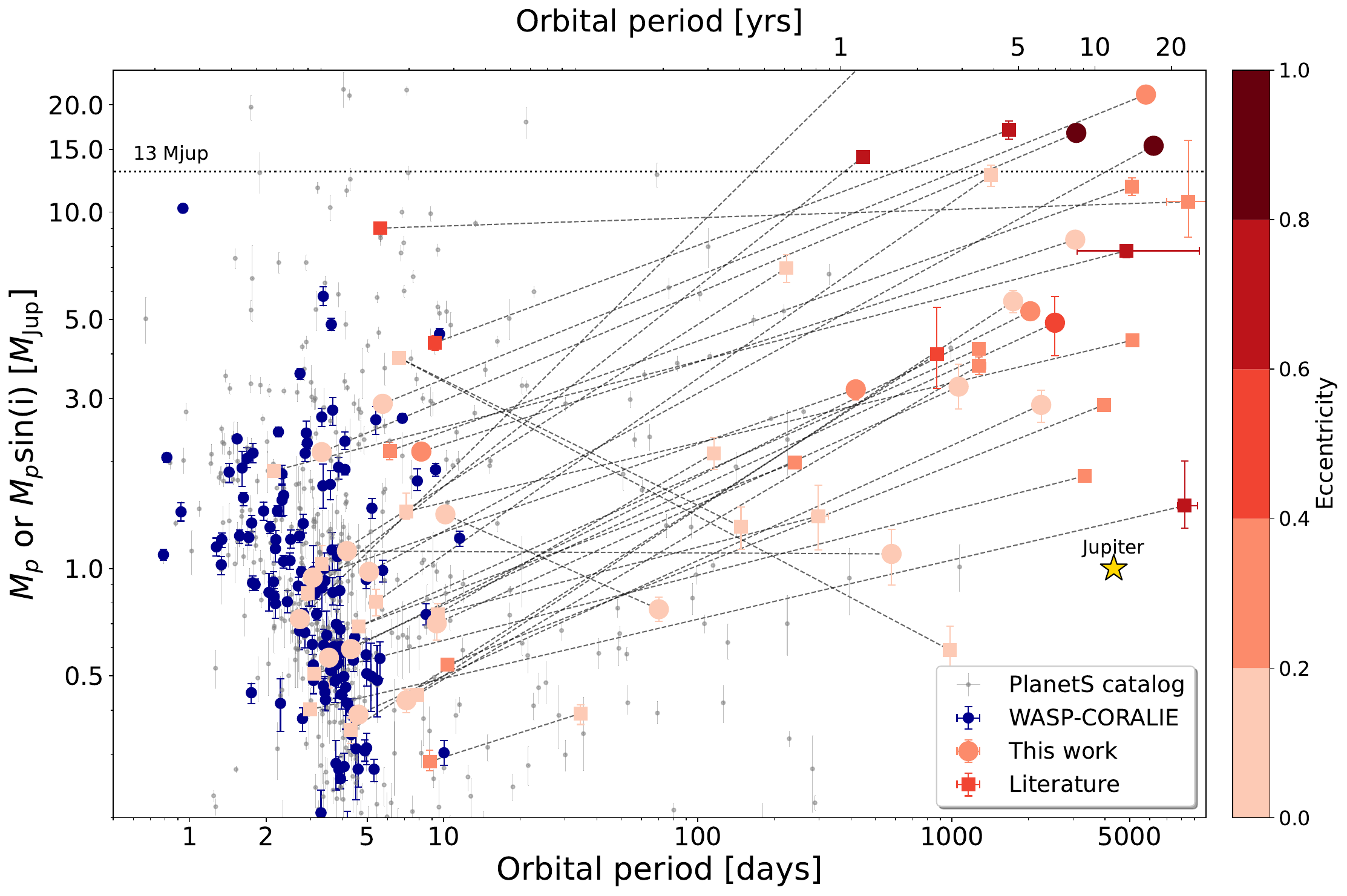}
\end{minipage}\hfill
\begin{minipage}[c]{5.5cm}
  \captionof{figure}{Same as Fig.~\ref{fig:mass_period_CORALIE}, but including the full combined sample of 30 hot Jupiters with detected outer companions from this study and the literature, used in the correlation analysis of Sect.~\ref{sect:discussion}.}
  \label{fig:mass_period_all}
\end{minipage}
\end{figure*}

\section{\textit{Gaia} detectability}\label{appendix:gaia}

For each system, we estimate the amplitude of the astrometric signal that the companion would induce on its host star, following Eq.~1--2 of \citet{Lammers2026}:

\begin{equation}
    \alpha \approx 95.4~\mu\mathrm{as} \times \frac{M_p}{M_J} \times \frac{a}{\mathrm{AU}} \times \left(\frac{M_\star}{M_\odot}\right)^{-1} \times \left(\frac{d}{10~\mathrm{pc}}\right)^{-1}
\end{equation}

Since the RV data only provide $M_p \sin i$, we use it directly in place of $M_p$, which gives a conservative lower limit on $\alpha$ and thus on the detectability of each system.

We then compare this expected signal to the precision \textit{Gaia} can achieve for a star of a given magnitude. Following Eq.~9 of \citet{Lammers2026}, the typical measurement uncertainty per \textit{Gaia} observation is:

\begin{equation}
    \sigma_{\mathrm{fov}}(G) = 54~\mu\mathrm{as} \times \max\left[1,\ 10^{0.2(G-14)}\right]
\end{equation}

A companion is considered detectable when the ratio of the astrometric signal to this measurement uncertainty, $\mathrm{SNR}_1 = \alpha / \sigma_{\mathrm{fov}}(G)$ (Eq.~7 of \citealt{Lammers2026}), exceeds 1.5 for \textit{Gaia} DR4 or 1.0 for DR5. These thresholds were calibrated by \citet{Lammers2026} using injection-recovery simulations.

Finally, for companions with orbital periods comparable to or longer than the time span covered by \textit{Gaia} (5.5~yr for DR4, 10.5~yr for DR5), we report the fraction of the orbit covered by the observations. A low fraction would normally make it difficult to fit a full Keplerian orbit from \textit{Gaia} astrometry alone. However, in our case the orbital period, eccentricity, and time of periastron are already well determined from the RV data, so \textit{Gaia} only needs to detect a deviation from single-star motion rather than recover the full orbit unassisted. Partial phase coverage is therefore less limiting here than in a blind astrometric search. This is summarized in the following Table~\ref{tab:gaia_detectability_confirmed}.

\begin{table*}
\centering
\caption{Astrometric detectability of the detected signals with Gaia DR4/DR5, following \citet{Lammers2026}.}
\label{tab:gaia_detectability_confirmed}
\begin{tabular}{lccccccccc}
\hline\hline
 & $P$ [yr] & $M_p\sin i$ [$M_\textrm{Jup}$] & $d$ [pc] & $\alpha$ [$\mu$as] & SNR$_1$ & Det. DR4 & Cov. DR4 [\%] & Det. DR5 & Cov. DR5 [\%] \\
\hline
\multicolumn{9}{l}{\textit{Detected companions}} \\
WASP-8 c  &  8.36 &  8.37 & 90 & 358.6 &  6.64 & Y & 65.8  & Y & 100.0 \\
WASP-22 c &  6.96 &  4.89 & 301 & 53.0 &  0.98 & N & 79.1  & N & 100.0 \\
WASP-41 c &  1.14 &  3.18 & 162 & 21.5 &  0.40 & N & 100.0 & N & 100.0 \\
WASP-47 c &  1.58 &  1.10 & 266 & 5.3 &  0.10 & N & 100.0 & N & 100.0 \\
WASP-53 c &  8.44 & 16.68 & 200 & 370.0 &  6.85 & Y & 65.1  & Y & 100.0 \\
WASP-67 c &  6.14 &  2.88 & 188 & 53.7 &  0.99 & N & 89.5  & N & 100.0 \\
WASP-68 c & 17.01 & 15.35 & 230 & 364.8 &  6.76 & Y & 32.3  & Y & 61.7  \\
WASP-81 c &  3.58 & 55.56 & 399 & 294.9 &  5.46 & Y & 100.0 & Y & 100.0 \\
WASP-99 c & 15.87 & 21.38 & 159 & 621.9 & 11.52 & Y & 34.7  & Y & 66.2  \\
WASP-132 c &  4.77 &  5.62 & 124 & 143.8 &  2.66 & Y & 100.0 & Y & 100.0 \\
WASP-134 c &  0.19 &  0.77 & 192 &  1.2 &  0.02 & N & 100.0 & N & 100.0 \\
WASP-171 c & 11.91 & 12.87 & 705 & 81.8 & 1.52 & Y & 46.2 & Y & 88.1 \\
WASP-185 c &  2.91 &  3.24 & 278 & 21.0 &  0.39 & N & 100.0 & N & 100.0 \\
\hline
\multicolumn{9}{l}{\textit{1 Keplerian model}} \\
WASP-34 c &  5.57 &  5.27 & 131 & 120.2 &  2.23 & Y & 98.8  & Y & 100.0 \\
\multicolumn{9}{l}{\textit{2 Keplerians model}} \\
WASP-34 c & 5.63 & 4.58 & 131 & 105.3 & 1.95 & Y & 97.6 & Y & 100.0 \\
WASP-34 d & 8.44 & 1.24 & 131 & 37.3 & 0.69 & N & 65.2 & N & 100.0 \\
\hline
\multicolumn{9}{l}{\textit{Ambiguous signals}} \\
WASP-7  & 3.29 & 2.09 & 161 & 23.3 & 0.43 & N & 100.0 & N & 100.0 \\
WASP-79 & 0.58 & 1.24 & 244 & 2.7 & 0.05 & N & 100.0 & N & 100.0 \\
WASP-84 & 1.36 & 0.63 & 100 & 8.2 & 0.15 & N & 100.0 & N & 100.0 \\
WASP-89 & 5.52 & 5.28 & 284 & 58.6 & 1.09 & N & 99.6  & Y & 100.0 \\
\hline\hline
\end{tabular}
\end{table*}

\end{appendix}

\end{document}